\documentclass[a4paper,fleqn,usenatbib]{mnras}

\usepackage{newtxtext,newtxmath}

\usepackage[T1]{fontenc}
\usepackage{ae,aecompl}
\usepackage{graphicx}

\usepackage{graphicx}	
\usepackage{soul}

\title[Superflares on Recurrent Nova V2487 Oph]{Discovery of Extreme, Roughly-Daily Superflares on the Recurrent Nova V2487 Oph}

\author[Schaefer, Pagnotta, \& Zoppelt]{
Bradley E. Schaefer$^{1}$\thanks{E-mail: schaefer@lsu.edu}  
Ashley Pagnotta$^{2}$ \&
Seth Zoppelt$^{2}$
\\
$^{1}$Department of Physics and Astronomy, Louisiana State University, Baton Rouge, Louisiana, 70820, USA\\
$^{2}$Department of Physics and Astronomy, College of Charleston, Charleston, SC 29424, USA\\
}

\pubyear{2021}

\begin{document}
\label{firstpage}
\pagerange{\pageref{firstpage}--\pageref{lastpage}}
\maketitle

\begin{abstract}

V2487 Oph is a recurrent nova with detected eruptions in 1900 and 1998.  Startlingly, V2487 Oph shows flares, called `Superflares', with up to 1.10 mag amplitude, fast rises of under one-minute, always with an initial impulsive spike followed by a roughly-exponential tail, typically one-hour durations, and with random event times averaging once-per-day.  The typical flare energy $E$ is over 10$^{38}$ ergs, while the yearly energy budget is 10$^{41}$ ergs.  V2487 Oph Superflares obey three relations; the number distribution of flare energies scales as $E^{-2.34\pm0.35}$, the waiting time from one flare to the next is proportional to $E$ of the first event, and flare durations scale as $E^{0.44\pm0.03}$.  Scenarios involving gravitational energy and nuclear energy fail to satisfy the three relations.  The magnetic energy scenario, however, can explain all three relations.  This scenario has magnetic field lines above the disc being twisted and amplified by the motions of their footprints, with magnetic reconnection releasing energy that comes out as Superflare light.  This exact mechanism is already well known to occur in white light solar flares, in ordinary M-type flare stars, and in the many Superflare stars observed all across the H-R diagram.  Superflares on Superflare stars have rise times, light curve shapes and durations that are very similar to those on V2487 Oph.  So we conclude that the V2487 Oph Superflares are caused by large-scale magnetic reconnection.  V2487 Oph is now the most extreme Superflare star, exhibiting the largest known flare energy (1.6$\times$10$^{39}$ ergs) and the fastest occurrence rate.

\end{abstract}




\section{Introduction}

V2487 Oph (Nova Ophiuchi 1998) was discovered as a very fast ($t_3$=8 days) nova on 1998 June 15, reaching a peak magnitude of 9.5 mag.  Infrared spectroscopy during the eruption showed  full-width half-maximum (FWHM) velocities of near 10,000 km s$^{-1}$ (Lynch et al. 2000), which is incredibly high for a nova event.  The only report of optical spectroscopy was two short sentences describing one spectrum by Filippenko et al. (1998).  The light curve displayed a plateau in brightness at $V$=13.7 from days 13--22, making it a P-class nova, just like most recurrent novae (RNe).  The eruption had a very short duration, lasting somewhat longer than 70 days, returning to quiescence with B=18.3 and V=17.7 (Schaefer 2010).

With the observational basis of the very fast rate of decline and the high ejection velocity, multiple groups suspected that V2487 Oph is an RN, for which only the latest eruption had been discovered (e.g., Hachisu et al. 2002).  To test these suggestions, we began a thorough search of many archival plates around the world going back to 1890.  We discovered a prior eruption in the year 1900 on a plate now stored at the Harvard College Observatory (Pagnotta et al. 2009).  This makes V2487 Oph the tenth known galactic RN.

The overall efficiency for discovery of eruptions from V2487 Oph is low.  The discovery efficiency for any one eruption before 1999 averaged to 1.2 per cent for undirected nova searches, while the directed archival searches had an average discovery probability of 30 per cent for any one eruption (Schaefer 2010).  With this, most of the eruptions between 1900 and 1998 were missed.  The best calculated recurrence time-scale is 18 years, so that 4 or 5 eruptions were missed between 1900 and 1999 (Pagnotta et al. 2009; Schaefer 2010).  A further implication of this is that V2487 Oph is likely to recur any year now.

V2487 Oph remains a relatively poorly observed system.  It was unknown before 1998, then it peaked at the magnitude of 9.5, followed by a very fast decline, so relatively few observations were made during eruption.  In quiescence, its magnitude was sufficiently faint that little attention was paid to the system.  In 2009, its RN nature was discovered, greatly raising the profile of V2487 Oph.  We began a long series of photometric observations, and we successfully proposed for {\it Kepler} to get 67 days of continuous photometry with 59 second time resolution during Campaign 9 of the {\it K2} Mission.  The biggest gap in observations is the complete lack of spectroscopy in quiescence.

In this paper, we report on our large set of photometry.  One of our original purposes was to determine the orbital period, $P$.  For this, we use three methods that point to a period close to 1.24 days.  Much more importantly, we made the startling discovery that V2487 Oph has frequent, large-amplitude flares.  Most of these flares have energies of $\gtrsim$10$^{38}$ ergs, and thus qualifying as Superflares.  These flares are sharply different from the usual flickering seen in all cataclysmic variables (CVs).  Indeed, nothing like these flares have ever been seen in any RN, classical nova, CV, X-ray binary, or contact binary in quiescence.  However, a wide variety of systems are known to have startlingly-energetic flares with similar properties in a number of other settings (Schaefer 1989; 1990; 1991; Schaefer, King, \& Deliyannis 2000), and these have been labelled as Superflares, with very wide-reaching implications (Rubenstein \& Schaefer 2000; Lingam \& Loeb 2017; Howard et al. 2018).  With this name, there is no necessary implication that the V2487 Oph Superflares share the same mechanism as the other classes of Superflares.  Still, the origin and nature of the unique Superflares are now the central mystery of V2487 Oph.

\section{Observations}

\begin{figure}
	\includegraphics[width=1.05\columnwidth]{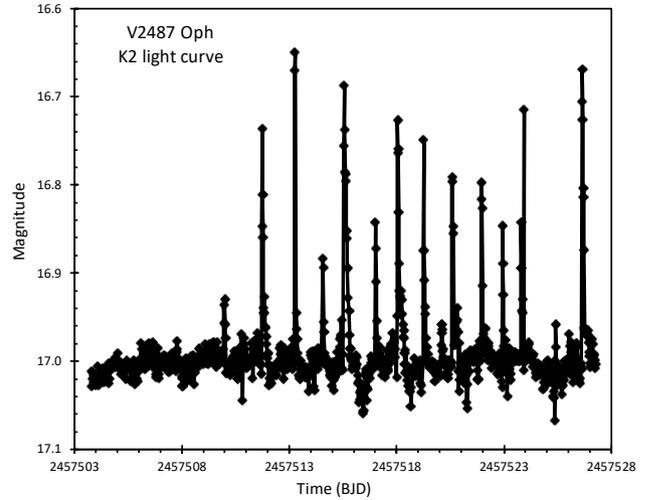}
    \caption{The first 25 days of the {\it K2} light curve with Superflares.  This light curve has 1800 second time resolution, with the Superflares only barely resolved, after the fluxes have been converted to a magnitude (with an arbitrary offset).  The Superflare amplitude is artificially low here, because the 1800 second time bins show the peaks (with maximum flux lasting only a few minutes) at a substantially lower value than is shown in the 59-second light curve.  This light curve shows that the flare frequency is not constant, with the first seven days having no large Superflares.  At first glance, the Superflares for this stretch of data (and only for this part of the light curve) appear to be uniformly spaced, but the uniformity is illusory with the time between peaks varying widely with no sharply preferred value.}  
\label{fig:lightcurve}    
\end{figure}

\begin{figure}
	\includegraphics[width=1.05\columnwidth]{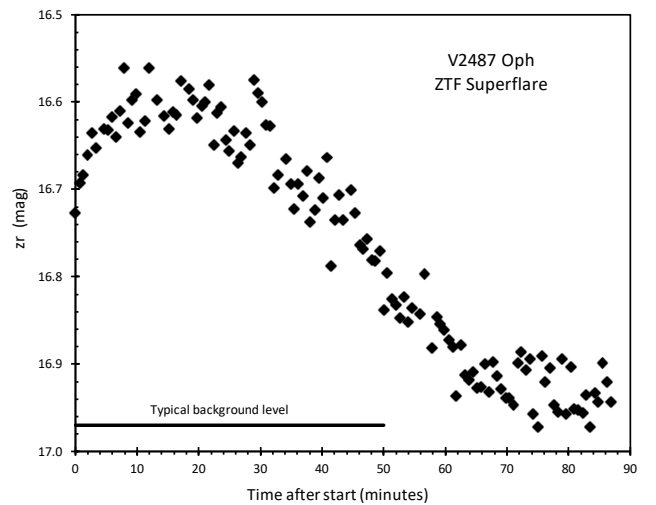}
    \caption{The Superflare recorded by ZTF.  This light curve provides the independent proof that V2487 Oph has Superflares, independent of any artefacts from the {\it K2} light curves.  Judging from the universal presence of sharp spikes at the start of Superflare light curves, it appears that the ZTF light curve started just a few minutes too late to catch the initial spike.}  
\label{fig:ztf_flare}    
\end{figure}

With a prior interest in RNe and the suggestions that V2487 Oph was a RN, we started a two-part study in 2002.  First, we looked for photometric modulation tied to the orbital period, as a long $P$ would further support the RN hypothesis.  This was carried out with CCD photometry on a variety of telescopes at McDonald Observatory and Cerro Tololo Inter-American Observatory (CTIO).  We started with just a few magnitudes per night, spread over many nights, but this revealed no periodic modulations, so we switched to intensive all-night monitoring for five straight nights.  No eclipses or sine wave variability were seen.  So when the {\it Kepler} spacecraft was used to look at the galactic centre region in Cycle 9 of the {\it K2} mission, we proposed taking 59-second integrations, and our proposal (GO 9912, PI Pagnotta) was accepted.  This resulted in an awesome photometric time series of high accuracy nearly continuously for 67 days (see Figure \ref{fig:lightcurve}).  From our initial study, we mistook the flaring to be the ubiquitous and highly-variable jumps due to thruster firings and pointing drift that occurred during all K2 observing \citep{2014PASP..126..398H}.  Only on the second detailed examination did we realise that the Superflares were unrelated to the thruster firings, and all other artefacts, so they must be intrinsic to V2487 Oph.  Looking at archived light curves, we found that the {\it Zwicky Transient Factory} (ZTF) happened to have a time series lasting 87 minutes (with 47 second time resolution) that beautifully shows a 0.35 mag amplitude Superflare with a FWHM of near 50 minutes and an exponential-like tail (see Figure \ref{fig:ztf_flare}).  This highly-significant flare provided the independent proof that V2487 Oph does indeed have Superflares.

The second part of our program was to test the RN hypothesis by seeking a prior nova eruption in the many archival sky photographs (plates) stored around the world.  Almost all the useful plates are now at the Harvard College Observatory\footnote{https://platestacks.cfa.harvard.edu/} and Sonneberg Observatory\footnote{https://www.astronomiemuseum.de/}, where we made multiple visits from 2004 to 2013 for the laborious task of examining all plates showing the field. In total, we examined 1960 plates at Harvard and 1800 plates at Sonneberg, that each could have shown a prior outburst (Pagnotta \& Schaefer 2014).  On one Harvard plate, from 1900 June 20, we found a $B$=10.27$\pm$0.11 image at exactly the position of V2487 Oph that shared the same unusual point spread function as all other stars on this plate caused by small trailing (Pagnotta et al. 2009). This was the confident discovery of the {\it second} nova event, making V2487 Oph the tenth known RN in our Milky Way.

\subsection{Photometry}

This paper uses large numbers of magnitudes measured for V2487 Oph.  This includes many of our own observations taken with the telescopes at McDonald Observatory and CTIO, plus our {\it Kepler} run in 2016, plus a variety of short light curves taken by various sky-surveys with on-line public availability.  Full details of the data sources are given in Table 1.  All the photometry from Table 1 is explicitly given in Table 2.  This table has 4698 line, all available as supplementary material, whereas only the first and last five lines are listed in the print version of this paper.  The 1800-second {\it K2} light curve is presented with the flux converted to a magnitude scale (normalized to an average of 17.000 mag).  The Pan-STARRS measures are given with the units of microJansky.  All known photometry {\it during eruption} is already tabulated in Schaefer (2010).

\begin{table*}
	\centering
	\caption{Photometric data for V2487 Oph}
	\begin{tabular}{lllllll} 
		\hline
		Telescope & Reference & Start Date  & JD start  &  JD end   &   Band    &   Number   \\
		\hline
McDonald 82-inch	&	This paper	&	2002 May 31	&	2452425	&	2452431	&	$BVRI$	&	38	\\
McDonald 0.8-m	&	This paper	&	2003 May 6	&	2452765	&	2452769	&	$V$	&	30	\\
AAVSO	&	[1]	&	2008 Aug 3	&	2454681	&	2459065	&	$BVRI$	&	148	\\
CTIO 1.3-m	&	This paper	&	2008 Sep 25	&	2454735	&	2454769	&	$BVRIJ$	&	41	\\
McDonald 0.8-m	&	This paper	&	2009 July 8	&	2455020	&	2455025	&	$V$	&	646	\\
Pan-STARRS	&	[2]; Chambers et al. (2016)	&	2010 Mar 25	&	2455280	&	2456756	&	$grizy$	&	52	\\
PTF	&	[3]; Rau et al. (2009)	&	2011 Mar 30	&	2455650	&	2455719	&	$g$	&	21	\\
{\it K2} Cycle 9	&	GO 9912, PI Pagnotta	&	2016 Apr 25	&	2457503	&	2457572	&	K2	&	2889	\\
ZTF	&	[4]; Bellm et al. (2019)	&	2018 Mar 28	&	2458205	&	2459458	&	$zg$, $zr$	&	833	\\
		\hline
	\end{tabular}
	\\
\begin{flushleft}
[1] https://www.aavso.org/data-download;~~
[2] https://catalogs.mast.stsci.edu/panstarrs/    \newline
[3] https://irsa.ipac.caltech.edu/cgi-bin/Gator/nph-scan?mission=irsa\&submit=Select\&projshort=PTF  \newline
[4] https://irsa.ipac.caltech.edu/cgi-bin/Gator/nph-scan?projshort=ZTF
\end{flushleft}
\end{table*}

\begin{table}
	\centering
	\caption{V2487 Oph magnitudes (full table available as supplementary material)}
	\begin{tabular}{lllll} 
		\hline
		HJD & Year & Band  & Magnitude  &  Source  \\
		\hline
2452425.8385	&	2002.412	&	V	&	17.368	$\pm$	0.050	&	McDonald 82"	\\
2452425.8394	&	2002.412	&	V	&	17.303	$\pm$	0.034	&	McDonald 82"	\\
2452425.8404	&	2002.412	&	V	&	17.282	$\pm$	0.028	&	McDonald 82"	\\
2452425.8432	&	2002.412	&	V	&	17.344	$\pm$	0.028	&	McDonald 82"	\\
2452425.8476	&	2002.412	&	V	&	17.333	$\pm$	0.027	&	McDonald 82"	\\
$\ldots$	&		&		&				&		\\
2459447.7133	&	2021.636	&	zr	&	16.965	$\pm$	0.028	&	ZTF	\\
2459450.6734	&	2021.645	&	zr	&	17.014	$\pm$	0.028	&	ZTF	\\
2459450.7122	&	2021.645	&	zg	&	17.043	$\pm$	0.044	&	ZTF	\\
2459454.6924	&	2021.656	&	zg	&	17.214	$\pm$	0.047	&	ZTF	\\
2459458.6911	&	2021.667	&	zr	&	17.134	$\pm$	0.029	&	ZTF	\\
		\hline
	\end{tabular}
\label{tab:all_mags}	
\end{table}

During the {\it K2} mission, with only two working gyroscopes for stability, the pointing of the camera suffered slight drifts, which were corrected by use of the spacecraft thrusters at roughly 6-hour intervals throughout the $\sim$67-day looks at target fields.  The slight drifts made each star image move slowly across the camera's chip, and the peak of the starlight would fall on different parts of pixels with differing sensitivities.  The result is a recorded flux that slowly brightens or fades across every six hours, only to be jerked back to the baseline when the thruster firings reset the pointing.  For V2487 Oph in Cycle 9, the effect appears as an accelerating loss of flux, even up to the 4 per cent level, only to have the light curve return to the baseline level every six-hours.  These jitters from the spacecraft pointing have light curves that have some similarities to the {\it rotated} light curves of the Superflares.  The corrections for any star cannot be known from first principles, because the intrapixel sensitivity variations cannot be mapped out effectively.  Instead, many workers developed methods to correct {\it K2} light curves, of which the primary method makes use of `co-trending basis vectors', derived from surrounding stars, and then fit to each light curve.  A full description of all the data, artefacts, and corrections is given in Kinemuchi et al. (2012).  The fully corrected light curves for the long-cadence data (with 1800 s time resolution) are generated and placed on the public website{\footnote{https://mast.stsci.edu/portal/Mashup/Clients/Mast/Portal.html}} of the Barbara A. Mikulski Archive for Space Telescopes (MAST), and in Table 2.

The 1800-s K2 light curve is good for many purposes, for example in looking for optical modulations on the expected orbital period of something like a day.  But 1800-s is comparable to the duration of the Superflares, which are poorly resolved.  To solve this, we have created a light curve with 59-second resolution.  Target pixel files (i.e., `postage stamps' showing the small area around the target) are also available with one image coming every 59-seconds.  For the 59-second K2 light curve, no pipeline corrected light curve is available for Campaign 9 data, and all methods provide imperfect corrections.  To this end, we have used the target pixel files and both the self flat-fielding corrector described in \citep{2014PASP..126..948V} and the EVEREST pipeline \citep{2016AJ....152..100L, 2018AJ....156...99L} to produce two 59-second resolution light curves. The differences between the two methods of correction do not appear to be significant. These light curves still have residual effects from thruster firings, and two of the Superflares have poor determinations of the post-flare backgrounds, leading to small shifts in the light curve shapes for the flare tails and small systematic errors in the fluences and peak fluxes.

The full 59-second {\it Kepler} light curve has 97400 magnitudes and is included as Table 3. Only the first and last five lines are presented in the print version of this paper, with the full 97400 lines available as supplementary material.  The table is simple, with just two columns, the first being the middle time of the 59-second integration expressed in Barycentric Julian Days (BJD).  This light curve still has uncorrected trends and residual effects from drifts.
    
For the flare light curves, which need full time resolution, we have separated out the Superflare light curves and explicitly list these in Table 4.  We have identified 60 Superflares in the {\it K2} light curve.  The light curves of these 60 flares are given with the non-flare flux subtraction, as based on a linear interpolation between pre-flare and post-flare time intervals.  Again, the print version displays only the first and last lines as examples, while the full 14570 lines are available as supplementary material.  In this table, the separate Superflares are identified with a running number (1--60), the second column is the mid-time of the integration in units of the spacecraft BJD which is fairly close to the more familiar HJD, and the third column is the background-subtracted {\it Kepler} flux in units of counts per second along with the one-sigma photometric errors.

\begin{table}
	\centering
	\caption{Full $K2$ light curve with 59-second time resolution (full table available as supplementary material)}
	\begin{tabular}{ll} 
		\hline
		Time (BJD)  & Flux (ct/s)  \\
		\hline
2457501.08718	&	2320	$\pm$	21	\\
2457501.08786	&	2339	$\pm$	21	\\
2457501.08854	&	2322	$\pm$	21	\\
2457501.08922	&	2345	$\pm$	21	\\
2457501.08990	&	2331	$\pm$	21	\\
...	&				\\
2457572.44379	&	1573	$\pm$	20	\\
2457572.44447	&	1535	$\pm$	20	\\
2457572.44515	&	1567	$\pm$	20	\\
2457572.44583	&	1584	$\pm$	20	\\
2457572.44651	&	1592	$\pm$	20	\\
		\hline
	\end{tabular}
\label{tab:flares}		
\end{table}

\begin{table}
	\centering
	\caption{Background-subtracted light curves for individual flares (full table available as supplementary material)}
	\begin{tabular}{llr} 
		\hline
		Flare & BJD& Flux (ct/s)  \\
		\hline
1	&	2457503.06262	&	7	$\pm$	20	\\
1	&	2457503.06330	&	1	$\pm$	20	\\
1	&	2457503.06398	&	11	$\pm$	20	\\
1	&	2457503.06466	&	-30	$\pm$	20	\\
1	&	2457503.06535	&	-11	$\pm$	20	\\
$\ldots$	&		&				\\
60	&	2457572.24830	&	3	$\pm$	20	\\
60	&	2457572.24898	&	-6	$\pm$	20	\\
60	&	2457572.24966	&	-55	$\pm$	20	\\
60	&	2457572.25034	&	-5	$\pm$	20	\\
60	&	2457572.25102	&	-25	$\pm$	20	\\
		\hline
	\end{tabular}
\label{tab:flares}		
\end{table}

\subsection{$K2$ Superflare Observations}

The Superflares are highly significant in terms of the Poisson measurement uncertainties.  However, they do have similarities to known artefacts in the {\it K2} photometry.  Indeed, when the {\it K2} light curve first came down, we initially mistook the Superflares for the effects of the slow positioning drift and the thruster firings.  As we are pointing to a previously unknown phenomenon, we must be very sure that the Superflares are not simply some instrumental artefact:  (1) The Superflares have light curve shapes that are generally fast-rise exponential-decline (FRED), whereas the thruster firing artefacts, visible in the raw count rates, are always an exponential-increasing fall with steepening slope followed by a jerk upwards, all in one or so bins.  The critical difference in shape demonstrates that the Superflares are not from thruster firings.  (2) The flare times are random with respect to the known times of thruster firings, with thruster events almost always being far from Superflares.  This breaks any possible connection.  (3) Detailed examination of the {\it K2} target pixel files shows no irregularities in the point spread function, for example, from passing asteroids or cosmic ray events.  (4) The excess flux from the flare light is exactly centroided at the position of the quiescent nova.  This severely limits the likelihood that the flares arise on any foreground star.  (5) The Superflare light curves appear the same in a variety of photometry apertures.  (6) Superflares do {\it not} appear in any nearby stars or in the background.  (7) The {\it Kepler} Team and many other researchers have all failed to detect any such artefacts or similar cases in the myriad of other stars with photometry of the same provenance.  (8) The well-resolved ZTF Superflare (see Fig 2) provides independent proof of the existence of Superflares on V2487 Oph.  (9) The ZTF light curve for the years 2018--2021 reveal ten additional Superflares.  In all, we have multiple proofs that the Superflares are real and intrinsic to V2487 Oph.

The flares are all isolated events, separate from each other, and with ordinary quiescence as the background state.  This is just one item which demonstrates a sharp distinction between Superflares and the ordinary CV flickering.  The shape of flickers in CV light curves is crudely triangular, with no fast rises or exponential declines.  CV flickering features on-going continuous burbling of flickers all run together, not the well-separated events seen in V2487 Oph.  The energy in Superflares is orders of magnitude larger than the energy in flickers.  With this, we do not expect that there is any similarity in physical mechanism between Superflares and flickering.  To avoid any confusion or merger of these two classes, a separate label is required.

The basic demographics are apparent from Figures 1 and 2.  The Superflares have a roughly once-per-day recurrence time, but their times are not periodic.  The typical amplitude is large fractions of a magnitude.  For the distance of $\sim$ 8000 pc and an absolute magnitude of +0.8 mag (see sections 3.3 and 3.5), a simple approximation of the peak luminosity would be $\sim$$10^{35}$ erg/sec.  The typical duration is around an hour.  This gives a luminous energy of order $10^{38}$ erg.  Flares this energetic are justifiably described as Superflares.

\begin{figure}
	\includegraphics[width=1.03\columnwidth]{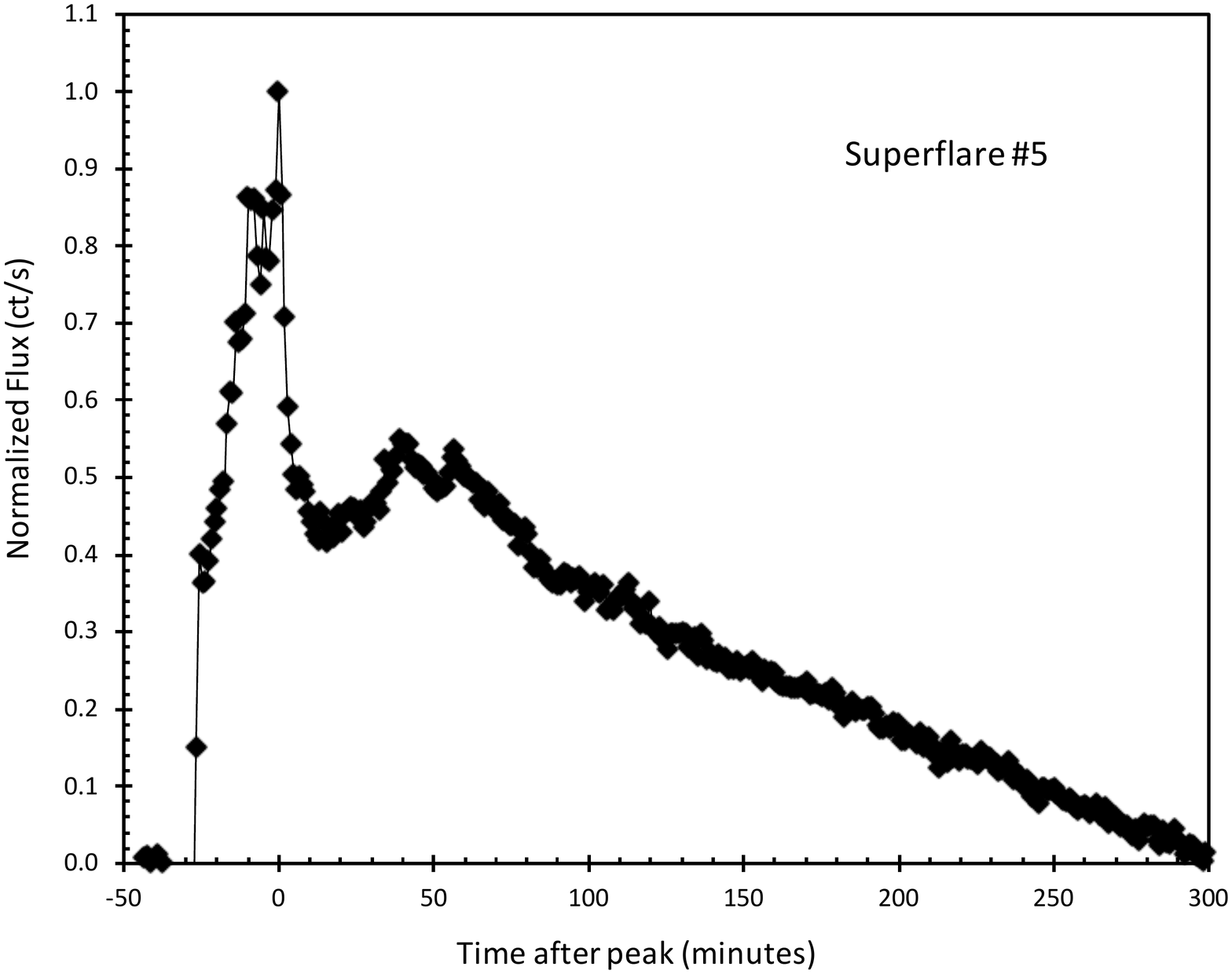}
	\includegraphics[width=1.03\columnwidth]{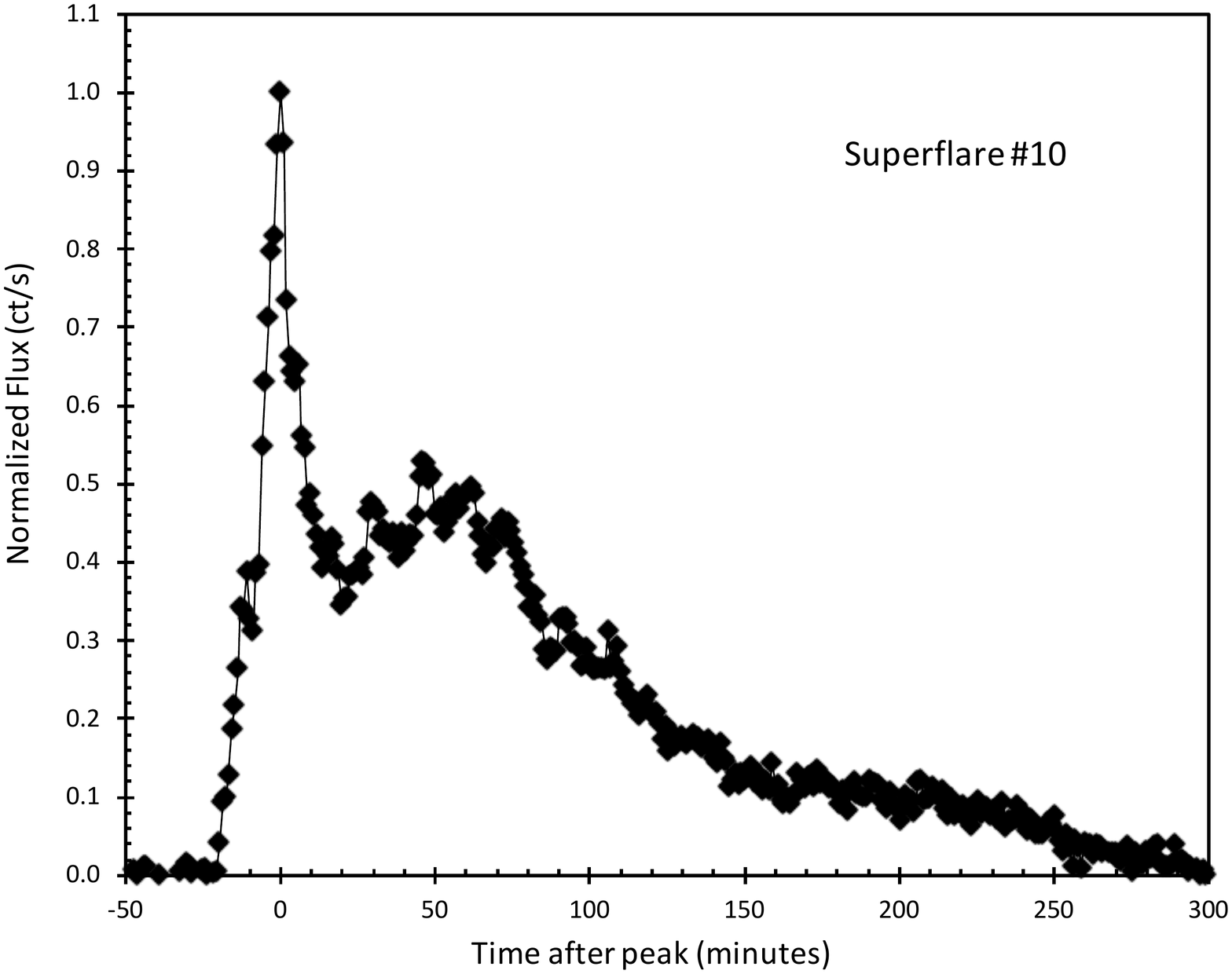}
	\includegraphics[width=1.03\columnwidth]{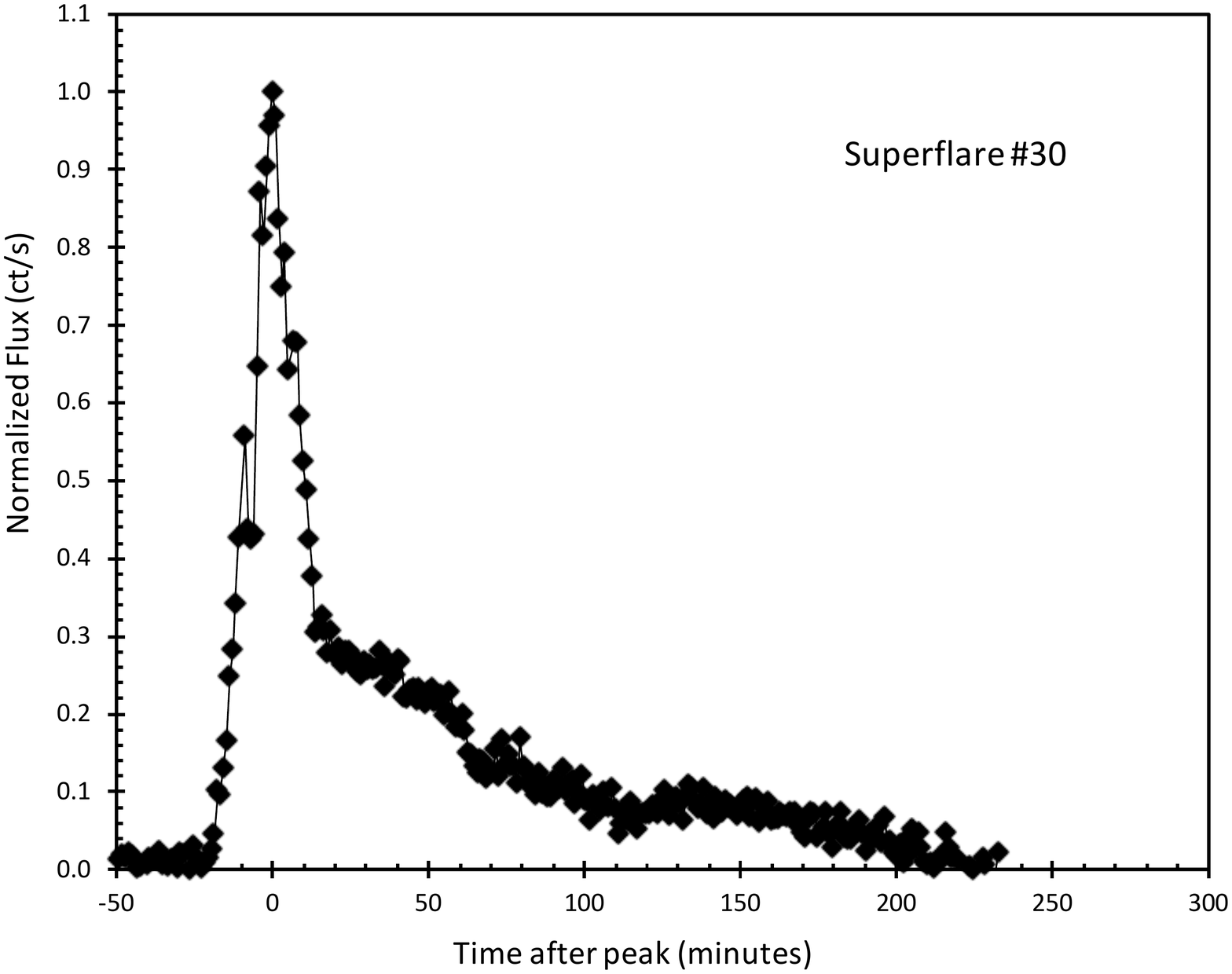}
  \caption{The three brightest Superflares.  These were chosen for having the brightest peak flux, so the relative photometric accuracy will be the best available.  (The photometric error bars are greatly smaller than each plotted symbol.)  These three light curves are typical for all 60 {\it K2} events.  The figures make apparent that all light curves consist of a sharp initial spike, plus a roughly-exponential tail that might have a rounded local peak near the start.  All the plots (including those in Fig 4) show the same time range for the background-subtracted fluxes normalized so that the peak is at 1.0.  }  
\end{figure}

Figure 3 displays the three brightest Superflare light curves, all with 59-second time resolution.  These were selected out solely so that we can show the light curves with the best available relative accuracy.  The Poisson error bars are greatly smaller than the plot symbols in all cases, so every small feature is highly significant and intrinsic to V2487 Oph.  Thus, the initial spikes (fast rises) for Superflares \#10 and \#30 look to have shorter and smaller spikes halfway up the primary rise, while \#5 appears to have the initial spike composed of two sharper spikes closely superposed on top of each other.  And the tops of the broad rounded bump near the start of the tail (the FRED peaks in \#5 and \#10) have significant several-minute-long bumps or flares superposed on top.  The fading tail of \#5 is closely linear for 220 minutes, so the description of the tails as `exponential declines' is valid only as a characterization of the decline as changing from a fast decline soon after peak to a slow decline as the background level is approached.

\begin{figure}
	\includegraphics[width=1.03\columnwidth]{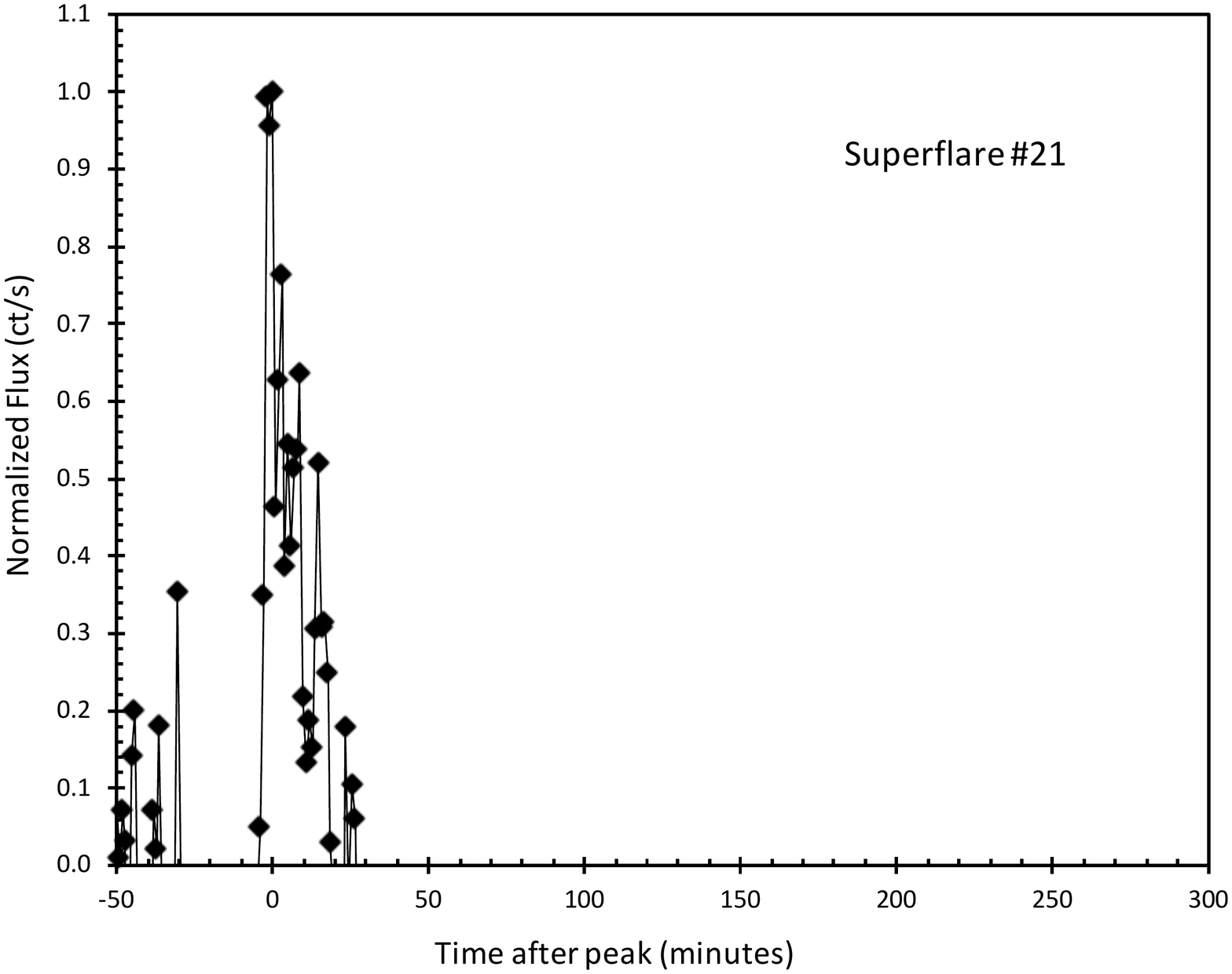}
	\includegraphics[width=1.03\columnwidth]{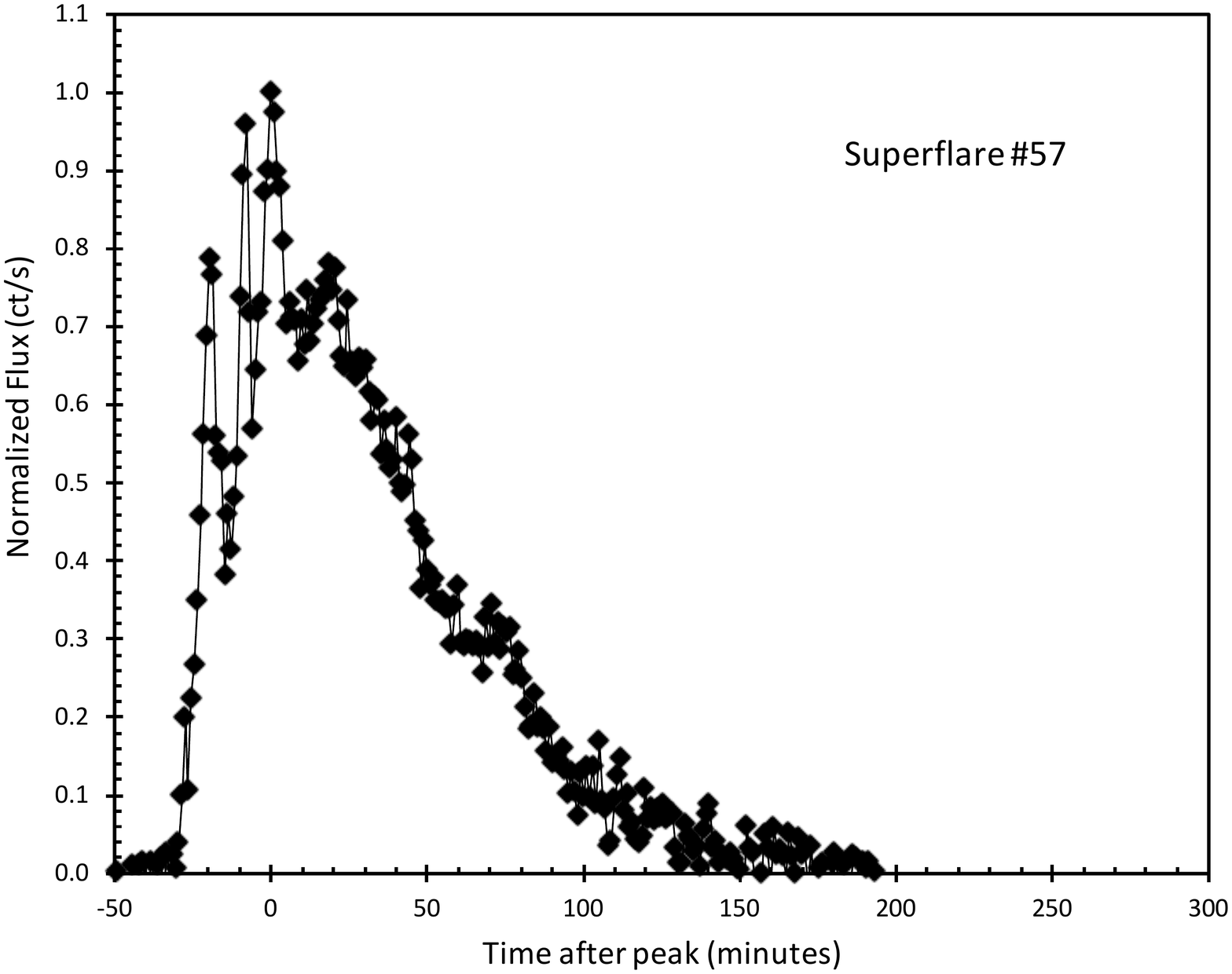}
	\includegraphics[width=1.03\columnwidth]{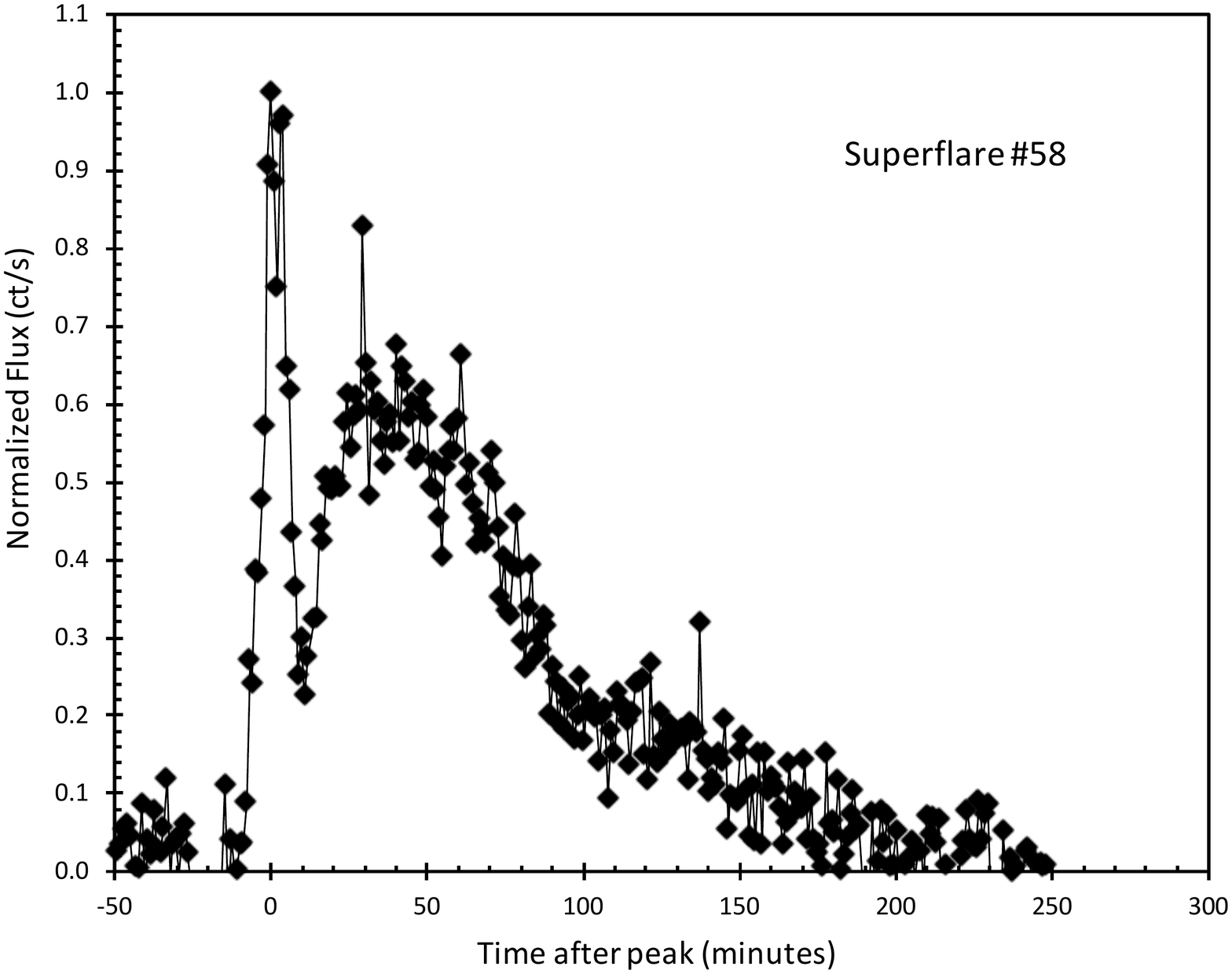}
  \caption{A sampling of three Superflare light curves.  These events were chosen to display the diversity of light curves.  Superflare \#21 is one of the faintest and shortest, showing the initial spike alone, presumably because the tail is too faint to be visible.  Superflare \#57 has three initial spikes near the start of a FRED.  Superflare \#58 looks like the ZTF flare (see Figure 2) except with the initial spike included in the coverage.  Photometric error bars are $\pm$0.17, $\pm$0.023, and $\pm$0.038 respectively for these faint flares.}  
\end{figure}

Figure 4 displays three additional Superflare light curves.  Figures 3 and 4 have all the same time resolution, range of time covered with respect to the peak, and are all normalized to the peak flux, so the shapes and durations of all six plots can be directly compared.  Between Figures 3 and 4, we have light curves that span the diversity of shapes. The striking feature of all the light curve shapes is that they all start off with some sort a short duration rapid-rise rapid-decline flare that we are calling a spike.  The spike durations range from one minute up to 30 minutes and usually correspond to the peak brightness of the flare.  The rise times are as short as $<$1 minute, and as long as 20 minutes.

The light curve shapes are very similar in composition.  All Superflares start off with at least one spike.  There are one-to-three initial spikes, which can have substantial overlap.  Following the spike, the light curve grades into either a FRED shape or a simple $\sim$exponential tail shape.  The relative amplitudes of the spike and the following emission varies substantially, sometimes with the spike being greatly brighter than all following emission, and sometimes the spike is only visible as a relatively small peak on the rising edge of the FRED.  The relative timing of the spike and the following component can vary substantially, sometimes with the spike superposed near the peak of the FRED and sometimes almost separate before the start of the FRED.  Sometimes the FRED has a rounded maximum, and sometimes the peak is quite sharp.  In some flares, it looks like a rounded FRED peak appears superposed on an exponential tail.  The tails are described as `exponential', only because they usually start with a relatively fast decline that flattens out as the light curve approaches the non-flare background level.  Despite this general description, the tails variously are closely linear at times, have bumps at other times, and have drop-offs at yet other times.  Throughout all this variability, we have the universal profile where Superflares start with a spike and are followed by various combinations of further spikes, FRED peaks, and `exponential' tails.

\begin{table*}
	\centering
	\caption{Superflares and their properties}
	\begin{tabular}{lrrrrrrrrrrrl} 
		\hline
Flare	&	Peak Time	&	$F_{max}$	&	Amp.	&	Fluence	&	Log[E]	&	$D_{eq}$	&	$T_{50}$	&	$T_{90}$	&	Rise	&	Fall	&	$\Delta T$	&	Flare light curve	\\
(\#)	&	(BJD)	&	(ct/s)	&	(mag)	&	(counts)	&	(erg)	&	(sec)	&	(sec)	&	(sec)	&	(sec)	&	(sec)	&	(days)	&	Shape	\\
		\hline
1	&	2457503.0919	&	1719	&	0.61	&	3881000	&	38.42	&	2258	&	3414	&	6533	&	412	&	3355	&	6.86	&	Spike+FRED	\\
2	&	2457509.9494	&	492	&	0.20	&	1930000	&	38.09	&	3925	&	3472	&	6886	&	59	&	1824	&	1.79	&	Spike+Spike+Tail	\\
3	&	2457511.7375	&	1599	&	0.53	&	5345000	&	38.51	&	3343	&	2648	&	5827	&	294	&	1471	&	1.52	&	Spike+Spike+Tail	\\
4	&	2457513.2614	&	1342	&	0.58	&	746000	&	37.79	&	556	&	824	&	1354	&	1001	&	1118	&	1.27	&	Spike+Spike+Tail	\\
5	&	2457514.5338	&	3009	&	1.10	&	17732000	&	39.21	&	5893	&	7004	&	14537	&	530	&	5885	&	0.82	&	Spike+FRED	\\
6	&	2457515.3492	&	206	&	0.09	&	133000	&	36.95	&	648	&	1118	&	2943	&	59	&	118	&	0.21	&	Spike	\\
7	&	2457515.5583	&	1809	&	0.75	&	17915000	&	39.19	&	9903	&	6945	&	14714	&	883	&	5120	&	0.64	&	Spike+FRED	\\
8	&	2457516.2020	&	148	&	0.07	&	171000	&	37.06	&	1152	&	942	&	2119	&	1354	&	530	&	0.82	&	Spike	\\
9	&	2457517.0222	&	1575	&	0.58	&	5409000	&	38.58	&	3434	&	5061	&	11830	&	235	&	4296	&	1.01	&	Spike+Spike+Tail	\\
10	&	2457518.0331	&	2819	&	0.91	&	12113000	&	38.95	&	4298	&	5650	&	13478	&	471	&	3649	&	1.23	&	Spike+FRED	\\
11	&	2457519.2592	&	1326	&	0.55	&	3727000	&	38.46	&	2810	&	3296	&	8534	&	235	&	2472	&	1.32	&	Spike+Tail	\\
12	&	2457520.5814	&	1437	&	0.53	&	6000000	&	38.61	&	4175	&	3002	&	8946	&	118	&	4708	&	1.06	&	Spike+Spike +Spike+Tail	\\
13	&	2457521.6447	&	215	&	0.10	&	198000	&	37.14	&	924	&	942	&	2472	&	1530	&	1118	&	0.30	&	Spike	\\
14	&	2457521.9431	&	1525	&	0.60	&	4341000	&	38.51	&	2846	&	3296	&	7063	&	235	&	1471	&	0.98	&	Spike+FRED	\\
15	&	2457522.9185	&	959	&	0.43	&	4021000	&	38.50	&	4191	&	4238	&	10241	&	235	&	3767	&	0.89	&	Spike+Spike+Tail	\\
16	&	2457523.8054	&	1821	&	0.65	&	3158000	&	38.34	&	1734	&	3060	&	6297	&	177	&	706	&	1.59	&	Spike+FRED	\\
17	&	2457525.3926	&	488	&	0.23	&	1443000	&	38.03	&	2954	&	2119	&	4885	&	59	&	2295	&	0.42	&	Spike+Spike+Tail	\\
18	&	2457525.8156	&	184	&	0.09	&	171000	&	37.11	&	925	&	1177	&	5003	&	59	&	59	&	0.21	&	Spike+Spike	\\
19	&	2457526.0275	&	307	&	0.15	&	417000	&	37.51	&	1359	&	1530	&	4061	&	59	&	1118	&	0.10	&	Spike+Spike+Tail	\\
20	&	2457526.1256	&	201	&	0.10	&	349000	&	37.43	&	1739	&	1766	&	4355	&	59	&	3767	&	0.12	&	Spike+FRED	\\
21	&	2457526.2489	&	120	&	0.06	&	28000	&	36.34	&	235	&	412	&	4120	&	59	&	118	&	0.07	&	Spike	\\
22	&	2457526.3224	&	198	&	0.10	&	146000	&	37.06	&	741	&	1118	&	2178	&	59	&	765	&	0.07	&	Spike+Spike	\\
23	&	2457526.3953	&	246	&	0.12	&	340000	&	37.42	&	1383	&	1648	&	3649	&	59	&	1766	&	0.23	&	Spike+FRED	\\
24	&	2457526.6269	&	1601	&	0.64	&	7905000	&	38.79	&	4938	&	4708	&	8710	&	118	&	4767	&	$\ldots$	&	Spike+Spike+FRED+FRED	\\
25	&	2457532.7214	&	728	&	0.35	&	2088000	&	38.23	&	2867	&	2648	&	7062	&	118	&	3472	&	0.66	&	Spike+Spike+Tail	\\
26	&	2457533.3794	&	263	&	0.15	&	1447000	&	38.10	&	5510	&	4649	&	9417	&	59	&	2236	&	0.82	&	Spike+FRED	\\
27	&	2457534.2002	&	854	&	0.43	&	2817000	&	38.39	&	3298	&	2295	&	5415	&	235	&	2472	&	1.68	&	Spike+FRED+Spike+Spike	\\
28	&	2457535.8807	&	655	&	0.30	&	1896000	&	38.16	&	2894	&	3355	&	7592	&	177	&	2236	&	1.06	&	Spike+FRED	\\
29	&	2457536.9393	&	1463	&	0.67	&	1722000	&	38.19	&	1176	&	883	&	3708	&	530	&	2060	&	1.16	&	Spike+Spike	\\
30	&	2457538.0980	&	2003	&	0.78	&	4888000	&	38.60	&	2441	&	4355	&	10535	&	235	&	1883	&	1.38	&	Spike+Tail	\\
31	&	2457539.4787	&	1724	&	0.75	&	4386000	&	38.59	&	2545	&	2884	&	7533	&	235	&	1707	&	2.60	&	Spike+Spike+Spike+FRED	\\
32	&	2457542.0767	&	996	&	0.45	&	3629000	&	38.47	&	3643	&	3237	&	7710	&	177	&	2060	&	3.89	&	Spike+FRED	\\
33	&	2457545.9717	&	1460	&	0.69	&	3377000	&	38.51	&	2313	&	3002	&	8475	&	1001	&	2472	&	2.80	&	Spike+Spike+FRED	\\
34	&	2457548.7672	&	1602	&	0.77	&	5888000	&	38.77	&	3676	&	4061	&	9652	&	647	&	2825	&	1.65	&	Spike+Tail	\\
35	&	2457550.4157	&	1058	&	0.53	&	5513000	&	38.70	&	5209	&	6592	&	13536	&	530	&	9946	&	1.29	&	Spike+Spike+FRED +FRED	\\
36	&	2457551.7031	&	229	&	0.13	&	353000	&	37.49	&	1542	&	1766	&	5120	&	59	&	1648	&	0.14	&	Spike+FRED+Spike+FRED	\\
37	&	2457551.8454	&	1365	&	0.64	&	6112000	&	38.74	&	4477	&	3178	&	6062	&	530	&	$\ldots$	&	0.89	&	Spike+FRED	\\
38	&	2457552.7316	&	1031	&	0.48	&	5196000	&	38.63	&	5038	&	3531	&	9416	&	177	&	2472	&	0.99	&	Spike+Spike+Spike+Tail	\\
39	&	2457553.7207	&	478	&	0.33	&	1892000	&	38.34	&	3960	&	4237	&	10064	&	294	&	5532	&	0.46	&	Spike+Spike+Tail	\\
40	&	2457554.1798	&	369	&	0.23	&	1089000	&	38.03	&	2950	&	4061	&	8004	&	177	&	5591	&	0.40	&	Spike+FRED	\\
41	&	2457554.5790	&	955	&	0.49	&	1918000	&	38.25	&	2008	&	1589	&	2884	&	177	&	$\ldots$	&	0.63	&	Spike+FRED	\\
42	&	2457555.2084	&	1012	&	0.55	&	7698000	&	38.89	&	7610	&	8239	&	21599	&	177	&	6121	&	0.96	&	Spike+Spike+FRED	\\
43	&	2457556.1702	&	1677	&	0.79	&	7252000	&	38.86	&	4323	&	6415	&	14654	&	235	&	6650	&	0.69	&	Spike+Tail+FRED	\\
44	&	2457556.8588	&	259	&	0.18	&	1076000	&	38.08	&	4155	&	3060	&	7769	&	59	&	4591	&	0.29	&	Spike+FRED	\\
45	&	2457557.1511	&	1463	&	0.73	&	4005000	&	38.61	&	2737	&	2884	&	7298	&	235	&	1883	&	0.71	&	Spike+Tail	\\
46	&	2457557.8622	&	1174	&	0.70	&	5674000	&	38.83	&	4834	&	4473	&	10947	&	177	&	1942	&	0.90	&	Spike+FRED	\\
47	&	2457558.7627	&	2041	&	0.88	&	3478000	&	38.52	&	1704	&	1295	&	3708	&	647	&	294	&	0.83	&	Spike+FRED	\\
48	&	2457559.5944	&	1637	&	0.78	&	7198000	&	38.86	&	4398	&	7533	&	18362	&	706	&	2707	&	0.69	&	Spike+Tail	\\
49	&	2457560.2871	&	1241	&	0.61	&	4497000	&	38.63	&	3625	&	4944	&	11771	&	824	&	3237	&	0.56	&	Spike+Spike+Spike+Tail	\\
50	&	2457560.8484	&	476	&	0.34	&	2253000	&	38.44	&	4735	&	3884	&	9063	&	471	&	2766	&	0.49	&	Spike+Spike+Spike+Tail	\\
51	&	2457561.3368	&	1381	&	0.86	&	3747000	&	38.71	&	2712	&	3355	&	10358	&	647	&	1942	&	1.39	&	Spike+Tail	\\
52	&	2457562.7250	&	991	&	0.58	&	3335000	&	38.57	&	3367	&	2531	&	6945	&	118	&	3531	&	1.50	&	Spike+Spike+Tail	\\
53	&	2457564.2270	&	829	&	0.60	&	2839000	&	38.60	&	3425	&	3767	&	10593	&	647	&	2648	&	0.79	&	Spike+FRED	\\
54	&	2457565.0192	&	157	&	0.11	&	56000	&	36.77	&	359	&	177	&	4473	&	59	&	177	&	0.20	&	Spike	\\
55	&	2457565.2228	&	1284	&	0.75	&	5279000	&	38.81	&	4110	&	5297	&	14831	&	765	&	2001	&	1.30	&	Spike+Spike+FRED	\\
56	&	2457566.5205	&	444	&	0.35	&	1698000	&	38.36	&	3829	&	3119	&	5944	&	177	&	1236	&	1.68	&	Spike+Spike	\\
57	&	2457568.2016	&	873	&	0.62	&	3425000	&	38.68	&	3924	&	3001	&	7357	&	177	&	2119	&	1.36	&	Spike+Spike+Spike+FRED	\\
58	&	2457569.5632	&	518	&	0.44	&	1971000	&	38.47	&	3808	&	3825	&	9534	&	294	&	4002	&	1.59	&	Spike+FRED	\\
59	&	2457571.1557	&	1071	&	0.78	&	2513000	&	38.58	&	2347	&	3413	&	6945	&	412	&	3355	&	1.06	&	Spike+Tail	\\
60	&	\underline{2457572.2142}	&	\underline{306}	&	\underline{0.27}	&	\underline{412000}	&	\underline{37.78}	&	\underline{1345}	&	\underline{942}	&	\underline{3355}	&	\underline{235}	&	\underline{1236}	&	\underline{$\ldots$}	&	Spike+Spike	\\
	&	Minimum =	&	120	&	0.06	&	28000	&	36.34	&	235	&	177	&	1354	&	59	&	59	&	0.07	&		\\
	&	Average =	&	1024	&	0.48	&	3671000	&	38.25	&	3154	&	3298	&	7974	&	346	&	2712	&	1.09	&		\\
	&	Maximum =	&	3009	&	1.10	&	17915000	&	39.21	&	9903	&	8239	&	21599	&	1530	&	9946	&	6.86	&		\\
		\hline
	\end{tabular}
\end{table*}

We have identified 60 Superflares in the {\it K2} light curve.  These are listed, along with their measured properties, in Table 5.  The minimum, average, and maximum values for many properties are tabulated at the bottom.  The first column is a running number identifying each Superflare.  The second column is the time of peak, in BJD.  The third column is the peak flux rate ($F_{max}$ in counts per second) above the interpolated background.  The fourth column is the amplitude of the flare in magnitudes, where the flare peak flux and the background flux are converted to a difference in magnitude.  The amplitudes are distributed with most being low and near the threshold for identifying flares, while a few Superflares are much brighter.  The brightest flare (\#5, see top panel of Fig 3) had a surprising amplitude of 1.10 mag.

The Superflare energy can be measured from the `fluence', which is the integral of the background-subtracted light curve over the entire duration of the event.  The fluence for each event is given in the fifth column of Table 5, with units of counts.  These counts will be proportional to the radiant energy of the flare, $E$.  For this calculation, we started with the derived absolute magnitude in quiescence of +0.8 (see Section 3.5), and scale by our Sun's absolute magnitude and luminosity, to get that V2487 Oph has an average luminosity in quiescence of 1.6$\times$10$^{35}$ erg/s.  Then, with the tabulated flare amplitude and the usual magnitude equation, we get the peak luminosity of the flare light.  This is then multiplied by the equivalent duration ($D_{eq}$), over which a constant luminosity would produce the same number of counts.  The result is the total energy, $E$, in units of ergs.  This calculation has two reasonable approximations; that the bolometric correction for the flare light is assumed to be the same as for our Sun's light, and that the quiescent level for each burst has $M_V$ of +0.8 mag.  Table 5 tabulates the logarithm of $E$ in ergs.  

We see that the typical Superflare has over 10$^{38}$ erg of energy, with a range of up to 1.6$\times$10$^{39}$ erg.  This is a tremendous amount of energy for one flare.  For example, the most energetic solar flares are weaker than 10$^{32}$ erg, while the typical Superflares on G-type main sequence stars are only up to 10$^{36}$ ergs (Maehara et al. 2012).  Ordinary main-sequence flare stars have energies only up to 10$^{35}$ ergs.  The total energy in our 60 observed flares is 2$\times$10$^{40}$ erg in 67 days of exposure, or over 10$^{41}$ ergs in a year.

Another important flare property is its duration.  With such irregular and variable light curves that asymptote to the background, there is no standard definition for duration.  So we have calculated three quantities that are like what we would call as a `duration'.  The first definition is for what we call the `equivalent duration', $D_{eq}$, which is the time over which the whole flare energy would come out at the peak flux level.  We can calculate $D_{eq}$ as the fluence divided by the peak flux.  The second and third definitions follow the usual procedure for gamma-ray burst durations, where $T_{50}$ and $T_{90}$ are the lengths of the time intervals that contain the central 50 per cent and 90 per cent of the total flare light.  That is, for example, the first quarter of the entire flare flux appears before the start of the $T_{50}$ interval, while the last quarter of the entire flare flux appears after the end of the $T_{50}$ interval.  These duration measures are sensitive to ordinary imperfections in the interpolation for estimating the quiescent flux level for subtraction.  The three durations, in units of seconds, appear in columns 7 to 9 of Table 5, and they differ from each other by factors of over 3$\times$ due to shape variations and noise in the flare light curves.

The rise and fall times of the light curve also have importance for understanding the physics of the Superflares.  Unfortunately, we know of no standard definition for either rise time or fall time, and any  algorithm to measure these times will give unexpected values due to the vagaries of the light curve shape plus the superposed `noise'.  After trying various definitions, we settled on reporting the minimum times for which the light curve either rises from 25 per cent to 50 per cent of peak flux, or falls from 50 per cent to 25 per cent of peak flux.  These values are tabulated in columns 10 and 11 of Table 5.

Another Superflare property of interest is the time between events, $\Delta T$.  This is tabulated in units of days in column 12 of Table 5.  The average flare-to-flare time interval is 1.09 days, justifying the characterization of these flares as being `daily'.  Nevertheless, there are substantial intervals with no Superflares, for example, there is a flare-less interval of 6.86 days near the start of the {\it K2} run (see Fig. 1).  The flare inter-arrival times are not clustered around some value (or perhaps including multiples of that value), so the flares are not periodic.  The distribution of flare inter-arrival times is roughly exponential (excepting the 6.86 day interval), which implies that the flare times are random and uncorrelated.

The last column in Table 5 contains a brief summary of the light curve components.  All light curves consist of various combinations of just three components, a narrow spike near the beginning, a long duration FRED shape with either a sharp or rounded peak, and an 'exponential' tail.

\subsection{$ZTF$ Superflare Observations}

The $ZTF$ light curve with 833 magnitudes covers the full observing seasons from 2018 to 2021.  Just over half of these magnitudes were taken on three nights with long time series.  One of these three nights resulted in the excellent observations of a Superflare, see Fig 2.  This one ZTF Superflare provides yet another proof that the Superflares are real and intrinsic to V2487 Oph, while not being any sort of artefact of the $K2$ observations.

The remainder of the magnitudes come from 125 nights, mostly with single observations.  Ten of the nights show a single bright magnitude with $zr$<16.6, which shows a Superflare.  The night HJD 2459309.0 has two back-to-back images with $zg$=17.02 and $zr$=16.47.  The flare-only light has a color of $zg$-$zr$=0.57, which is similar to the color of the underlying system.

These 11 ZTF flares show that the Superflare phenomenon is on-going from 2018--2021.  With 11-out-of-128 nights having superflares, the fraction of time well-above the background level is 9 per cent.  This is consistent with an average of once-per-day flaring (from the $K2$ light curve).  With on-going activity, this provides hope that observations in the 2022 observing season can catch Superflares.

\subsection{Spectral Energy Distribution}

The spectral energy distribution (SED) of V2487 Oph can be used to separate out the companion star flux from the disc component.  In particular, with the light curve similarity to U Sco, the weak expectation is that the companion is a $\sim$6000 K sub-giant star that might appear on top of the disc component.  Such could provide proof that V2487 Oph is with a sub-giant companion, and hence with a $\sim$1-day orbital period.  Identification of the companion flux in the SED could also give a blackbody distance to V2487 Oph.

We have constructed an SED from seven data sources, with the details in Table 6. Although the time of the observations is very widespread, so is the wavelength coverage. We have positive detections from the {\it GALEX} NUV band centred on 0.231 micron out to the second {\it WISE} band centred at 4.6 microns.  Corrections for extinction are critical, and we have used the corrections for $E(B-V)$=0.65 mag (see Section 3.4).  Details on the analysis were taken from Table 31 of Schaefer (2010), plus the mission documents for {\it GALEX} and {\it WISE}.  The result is a list of the the extinction corrected flux ($F_{\nu,0}$ in units of Janskys) versus the central photon frequency ($\nu$ in Hertz).  The SED for V2487 Oph in quiescence is plotted in Figure 5.  The measures from Lynch et al. (2000) are in the late tail of the eruption, so not included in this plot for the quiescent state of the nova.

\begin{table*}
	\centering
	\caption{Spectral Energy Distribution for V2487 Oph)}
	\begin{tabular}{lllll} 
		\hline
		Data Source & Band & Flux (unit)  & $\log[\nu]$ (Hz)  &  $\log[F_{\nu,0}]$ (Jy)   \\
		\hline
GALEX (J173159.7-191356)	&	$NUV$	&	20.29	 (AB mag)	&	15.11	&	-2.49	\\
Schaefer CTIO 1.3m, 25 Sep 2008	&	$B$	&	18.16	 (mag)	&	14.83	&	-2.58	\\
Schaefer CTIO 1.3m, 25 Sep 2008	&	$V$	&	17.44	 (mag)	&	14.74	&	-2.61	\\
Schaefer CTIO 1.3m, 25 Sep 2008	&	$R$	&	16.95	 (mag)	&	14.67	&	-2.69	\\
Schaefer CTIO 1.3m, 25 Sep 2008	&	$I$	&	16.34	 (mag)	&	14.58	&	-2.74	\\
Schaefer CTIO 1.3m, 25 Sep 2008	&	$J$	&	15.46	 (mag)	&	14.38	&	-2.75	\\
Pan-STARRS, 2010--2014	&	$g$	&	342	 ($\mu$Jy)	&	14.79	&	-2.52	\\
Pan-STARRS, 2010--2014	&	$r$	&	577	 ($\mu$Jy)	&	14.68	&	-2.54	\\
Pan-STARRS, 2010--2014	&	$i$	&	691	 ($\mu$Jy)	&	14.60	&	-2.61	\\
Pan-STARRS, 2010--2014	&	$z$	&	857	 ($\mu$Jy)	&	14.54	&	-2.67	\\
Pan-STARRS, 2010--2014	&	$y$	&	921	 ($\mu$Jy)	&	14.49	&	-2.71	\\
DENIS	&	$I$	&	16.25	 (mag)	&	14.58	&	-2.71	\\
DENIS	&	$J$	&	15.54	 (mag)	&	14.38	&	-2.78	\\
Lynch et al. (2000), Lick 3-m, 2 Oct 1998	&	0.8 $\mu$	&	6.0$\times$10$^{-19}$	 (W/cm/$\mu$)	&	14.57	&	-2.51	\\
Lynch et al. (2000), Lick 3-m, 2 Oct 1998	&	1.1 $\mu$	&	4.5$\times$10$^{-19}$	 (W/cm/$\mu$)	&	14.44	&	-2.46	\\
Lynch et al. (2000), Lick 3-m, 2 Oct 1998	&	1.3 $\mu$	&	2.6$\times$10$^{-19}$	 (W/cm/$\mu$)	&	14.36	&	-2.63	\\
Lynch et al. (2000), Lick 3-m, 2 Oct 1998	&	1.55 $\mu$	&	2.0$\times$10$^{-19}$	 (W/cm/$\mu$)	&	14.29	&	-2.65	\\
Lynch et al. (2000), Lick 3-m, 2 Oct 1998	&	1.8 $\mu$	&	1.2$\times$10$^{-19}$	 (W/cm/$\mu$)	&	14.22	&	-2.77	\\
Lynch et al. (2000), Lick 3-m, 2 Oct 1998	&	1.97 $\mu$	&	1.1$\times$10$^{-19}$	 (W/cm/$\mu$)	&	14.18	&	-2.77	\\
Lynch et al. (2000), Lick 3-m, 2 Oct 1998	&	2.2 $\mu$	&	1.0$\times$10$^{-19}$	 (W/cm/$\mu$)	&	14.13	&	-2.71	\\
2MASS (J17315980-1913561)	&	$J$	&	15.36	 (mag)	&	14.38	&	-2.71	\\
2MASS (J17315980-1913561)	&	$H$	&	14.88	 (mag)	&	14.27	&	-2.78	\\
2MASS (J17315980-1913561)	&	$K$	&	14.41	 (mag)	&	14.13	&	-2.85	\\
WISE (J173159.85-191355.6)	&	$W1$	&	14.13	 (Vega mag)	&	13.95	&	-3.10	\\
WISE (J173159.85-191355.6)	&	$W2$	&	14.44	 (Vega mag)	&	13.81	&	-3.49	\\
		\hline
	\end{tabular}
\end{table*}

\begin{figure}
	\includegraphics[width=1.05\columnwidth]{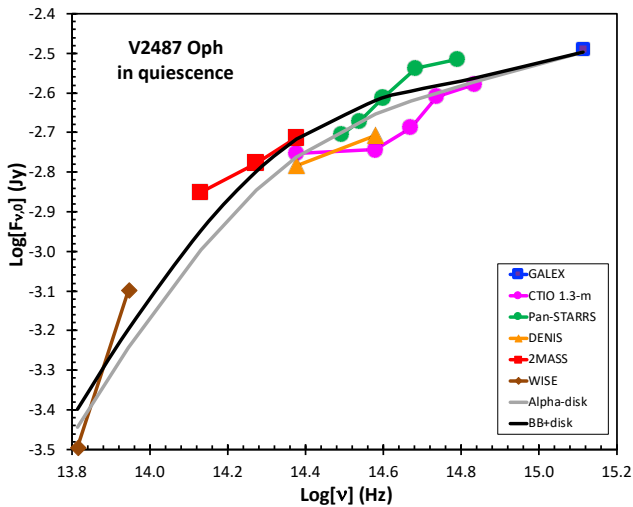}
    \caption{The spectral energy distribution of V2487 Oph.  The SED is dominated by the disc light, as no reasonable companion star can provide the NUV flux of {\it GALEX}.  The turn down going towards the infrared is the usual cut-off from the edge of a normal disc.  The best-fitting alpha-disc model is shown by the gray curve, with a best-fitting accretion rate of 5$\times$10$^{-8}$ M$_{\odot}$ yr$^{-1}$.  The frequency of this cutoff is controlled by the radius of the outer edge of the accretion disk, with the estimated radius of (3.2$\pm$0.4)$\times$10$^{11}$ cm.  This gives an orbital period of 1.1--2.5 days.  Any contribution from the blackbody of a companion star provides only minimal flux.  The black curve shows the maximum-acceptable contribution of a 6000 K blackbody added on to the best-fitting disc-only model.}  
\end{figure}

The SED shows a sharp rise from the infrared-to-red and a flattening off from the blue-to-ultraviolet.  At first glance, this looks like it might be a simple blackbody shape, but this is not so, as the curvature is much too broad to be a blackbody.  Further, for any reasonable temperature for the companion star, $T_{comp}$, there is no significant flux in the {\it GALEX} NUV band, so there must be a large disc component and it must dominate over much of the SED.  On the blue side of the SED, the slope is consistent with the expected $\nu ^{+1/3}$ spectrum of the middle of an accretion disc.  But the downturn to the left from the {\it GALEX} point shows that the SED cannot be the simplistic $\nu ^{+1/3}$ all the way to the infrared.  Rather, we have calculated the disc component by the full integration of the alpha-disc model flux for annuli stretching from near the white dwarf all the way out to the outer edge of the disc, as prescribed in Frank, King, \& Raine (2002) and many other places.   The companion star will contribute flux that is close to a single-temperature blackbody, although presumably a small fraction of its inward hemisphere will be irradiated to a somewhat higher temperature.  The measurement errors are all greatly smaller than the plot symbols, so the scatter is almost entirely due to the year-to-year variations in the quiescent flux.  The vertical scatter visible in the SED has an RMS of roughly 15 per cent. The companion star contribution to the SED cannot vary by much, so the observed variations must arise from the disc and a variable accretion rate.

We can briefly consider whether any other components contribute significantly to the SED.  The white dwarf itself is greatly too small to contribute any substantial flux in Fig 5 due to its small emitting area, no matter the temperature.  Any nova shell from the 1998 eruption can only be greatly too faint to matter, and such is not visible in the 2MASS images.  V2487 Oph likely has a magnetic white dwarf (Hernanz \& Sala 2002), so it might be an intermediate polar with an inner accretion column giving off cyclotron emission.  But for the intermediate polar case, $B$$<$10$^7$ G, so the cyclotron frequency corresponds to a wavelength of $>$11 $\mu$, which is to say it will not contribute in Fig 5.  The possibility of other components making substantial contributions to the SED is precluded by the good fit of the simple and expected alpha-disk model to the observed SED.  For example, the spectrum of a broad cyclotron line, even with harmonics, has no place to rise above the disc flux.  Even the expected blackbody spectrum from the companion star can at most provide only negligible light to the SED.  So largely, fits of spectral models to the SED only have the alpha-disk flux.

The task is then to fit the observed SED to a disc+blackbody model.  This task has three fundamental challenges.  The first problem is that the fluxes vary by 15 per cent from year-to-year (from intrinsic variability), with the various SED segments having some systematic offset from each other.  This is problematic for anything like a chi-square fit, because the real error bars are not the negligibly-small measurement errors, but rather are nearly 15 per cent errors arising from the intrinsic variations. The second problem is that the measures from Lick Observatory (Lynch et al. 2000) apparently have lingering flux from the 1998 eruption.  Their SED contribution is from 109 days after the eruption peak, while the V-band light curve shows V2487 Oph fast approaching the quiescence level even on day 70.  We judge that the somewhat brighter flux in the SED is likely due to the eruption.  If the data are taken to have contamination from the eruption, then the Lynch et al. points should not be used to evaluate the system in quiescence, and so we have not included them in the chi-square fits discussed below.  The third problem is that the slope between the two {\it WISE} points is steeper than possible for any companion blackbody or any alpha-disc.  These two points have a spectral slope corresponding to $\nu ^{+3.0}$, which is substantially steeper than a Rayleigh-Jeans slope with $\nu ^{+2}$.  So this means that something is wrong: while we do not think that it is observational error, the remaining possibilities include the fluxes being taken on different dates or line emission in the WISE1 band.  Despite the apparent presence of some sort of a problem with the WISE fluxes, we expect that the fluxes are approximately good, and that we should work forward.

We have tried chi-square fits to the SED (not including the Lick 3-m data as it is in the tail of the eruption) in Table 6, taken at face value.  We have adopted 15 per cent one-sigma error bars on all points, with this likely being appropriate for the nova's intrinsic variability, and this returns reduced chi-squares around unity.  Our series of SED fits were made with the disc and blackbody components having arbitrarily normalization, chosen only so as to optimize the fit, so we are only fitting {\it shapes} here.  To start, we ran models where only the disc was present.  In this case, the only parameters that can significantly change the model SED are the arbitrary normalization, the accretion rate, and the fraction of the white dwarf's Roche lobe filled by the disc.  The best-fittings are for when the accretion rate is close to $\dot{M}$=5$\times$10$^{-8}$ M$_{\odot}$ yr$^{-1}$, with the disk filling most of the Roche lobe.  (This very high accretion rate is as required for V2487 Oph to have a short recurrence time scale.)  This disc-only best-fitting is displayed in Fig. 5 as the gray curve.  In all, we have a satisfactory fit (with $\dot{M}$$\sim$5$\times$10$^{-8}$ M$_{\odot}$ yr$^{-1}$ filling the Roche lobe) for the disc alone.

The turn over to long wavelengths is due to the cut off at the outer edge of the disc.  That is, a large disc will have large outer cool areas contributing bright flux in the {\it WISE} bands, while a small disc will have no outer cool areas contributing IR radiation.  So the observed cut off, from {\it GALEX} to {\it WISE} is primarily a function of the disc radius.  From our fits to the SED with only an $\alpha$-disc, we derive that the outer edge of the accretion disc has a radius of (3.2$\pm$0.4)$\times$10$^{11}$ cm.  This sets the scale for the Roche lobe size and for ordinary assumptions about the stellar masses, then translates into an orbital period.  We expect that the accretion disc size will be of order 80 per cent of the white dwarf's Roche lobe size.  For the white dwarf with 1.35 M$_{\odot}$ (see Section 3.2) and the companion star of $\sim$1.0 M$_{\odot}$, our best estimate for the required orbital period is 2.0$\pm$0.4 days.  This derived period has negligible dependency on the companion mass over any plausible range.  If we take the white dwarf mass to be 1.25 solar masses, then the period is 2.1$\pm$0.4 days.  If we take the disc to fill the Roche lobe, the required period is 1.4$\pm$0.3 days.  Given the uncertainties in the SED, it is hard to put a formal error bar on the derived period, but anything in the range roughly from 1.1--2.5 days seems reasonable.  In particular, $P$=1.24 days seems good, provided the disc fills much of the Roche lobe.  In this solution, the companion star has a radius of 2.3 $R_{\odot}$, substantially larger than any plausible main sequence star.  This result is important because it set limits on the orbital period.  The $P$ cannot be shorter than 1.1 day nor longer than 2.5 days. This is a useful constraint for looking for orbital photometric periods.  This also shows that the companion is not a main sequence star and not a giant star.  The companion must be a sub-giant, with some evolution off the main sequence.

Next, we tried to add in a blackbody for the companion star.  That the disc-only fit works is telling us that the companion provides only minimal flux.  The satisfaction with the disc-only fit is also telling us that the companion star is no sort of red giant star (like in many RNe systems, including T CrB and RS Oph).  The deviations above the gray curve are showing us the flux that can be explained by a companion, and there is little excess flux.  To illustrate this, the black curve shows a $T_{comp}$=6000 K added to the gray-line disc model.  This disc+blackbody model has a higher chi-square than the disc-only model, yet it still looks acceptable given the substantial uncertainties in the SED.  We have found acceptable disc+blackbody fits for ranges of $T_{comp}$ from around 3000 to 7000 K.  The blackbody flux is a small fraction of the total flux. 

For use in later calculations, we should specify the properties of these blackbody fits.  For the case of the 6000K blackbody with the maximum acceptable flux, the companion provides only 22 per cent of the total flux in the V-band, with this flux equaling 0.00052 Jy.  At the peak of $F_{\nu,0}$ for the companion star, at a frequency of 3.5$\times$10$^{14}$ Hz, the flux is 0.00069 Jy.  For the case of the highest acceptable flux from a 3000 K blackbody, the companion only contributes 2.2 per cent of the V-band flux, which is 0.000049 Jy.  At the peak of $F_{\nu,0}$ for the companion star, at a frequency of 1.7$\times$10$^{14}$ Hz, the flux is 0.00065 Jy.  Given the acceptability of the zero-blackbody fit, these fluxes from the companion star can only be viewed as upper limits.

\section{Analysis For System Properties}

\subsection{Orbital Period}

The single most important property of CVs is the orbital period, so much of our work was aimed at finding a photometric modulation on the orbital period.  To this end, starting in 2002, we collected substantial photometry during various observing runs at CTIO, near La Serena, Chile, and at the McDonald Observatory in western Texas.  We were disappointed to find no eclipses and no roughly-sinusoidal modulations.  This can be generalized and quantified by running a discrete Fourier transform (DFT).  With this, we find no significant peaks from periods of 0.04 to 20 days.  When we look at the further observing runs of AAVSO, PTF, and ZTF, we still come up with no blatant peak in the DFT.  Well, there are many peaks visible in the various DFTs, but none is repeatable or significant above the level expected for ordinary flickering.  When DFTs are run for various combinations of observing runs, we just see a complex structure of aliases, with no peak standing out or repeatable.

A plausible try to overcome the natural strong beating present in DFTs of multiple widely-separated runs is to co-add the Fourier {\it power} of the individual runs.  This old technique was used to pull out the correct alias for the RN T Pyx from a large set of widely separated observing runs, all with no clear periodicity (Schaefer et al. 1992).  When tried for V2487 Oph, the co-addition of the individual DFTs for the many ground-based observing runs does show an apparently significant peak at 1.25$\pm$0.03 days.  That is, a period near 1.25 days is the only period that produces DFT peaks consistently in all the observing runs.  (By the nature of the problem, some very fine-tuning of the period will produce a good-enough phasing for all the observing runs.) Running a Lomb-Scargle Periodogram using the Lightkurve Python package returned a period of 1.23$\pm$0.02 days. Thus 1.24$\pm$0.02 days might well be the correct orbital period for V2487 Oph.

The {\it K2} light curve alone is of such impressive accuracy and coverage that any coherent photometric periodicity should be apparent.  When we run a DFT of the entire 67 days, a single peak at a period near 1.24 days does standout with twice as much power as any other peak.  This appears as a good independent confirmation of the 1.24$\pm$0.02 day periodicity from all ground-based runs.  In an effort to test the period, we split the light curve at its natural break point during the data gap around BJD 2457530.  To our dismay, this last two-third of the {\it K2} light curve has only a small DFT peak at a consistent period, indeed, it is only the second highest peak (and {\it K2} has no problems with aliases), apparently at a level that cannot be considered significant.  For the first third of the light curve (see Fig. 1), the single strong DFT peak appears at 1.2216$\pm$0.0031 days.  If we only look at the flare-less interval before BJD 2457509.8 in the 1800-s light curve, then we see no DFT power or peak anywhere near 1.25 days.  So almost all of the power for the peak comes from the short interval from BJD 2457509.8 to 2457527.3, with a period of 1.2286$\pm$0.0017 days.  On looking at the folded light curve, we see that the quiescent level has no significant modulation, while all the signal arises because the 23 Superflares roughly line up in phase.  On constructing a light curve  with the flares excised for the same time interval, the DFT shows no peak or any non-random amount of power anywhere near the 1.2286 day periodicity.  All this is saying that the apparent periodicity of $\sim$1.23 days is solely in the flare light.  

With the Superflares being of short duration, this can only mean that their {\it times} are quasi-periodic, and only for one 18-day interval.  Looking at the $\Delta T$ times between flares \#2 to \#24 in Table 5, we do not see even a vague clustering around 1.23 days.   The DFT peak would be caused mainly by the brightest flares happening to line up around the same phase.  In this case, the eight peaks with $\log [E]$$>$38.5 happen to cover only a range of 0.3 in phase space, and this is adequate to produce the observed peak.  Such DFT peaks will be formally significant because of the flares have a huge flux, and because the noise model for the significance calculation assumes uncorrelated Gaussian noise rather than flares.  Such a confluence of phases is expected for random flare times, when we can widely vary the trial periods and we can select flares from a wide range of time intervals.  That is, any random set of 60 flare times superposed on a flat background will always produce some falsely-significant DFT peaks for some period for some careful selection of flares.  In all, the $\sim$1.23 day periodicity is arising from flare times lining up in phase, and this could be due to the system producing flares with times tied to the orbit.

So what is the orbital period of V2487 Oph?  The widespread expectation has been that V2487 Oph is a sister-RN of U Sco, so the period is something around 1.23 days, but this expectation is greatly too weak to be useable.  Still, we have the same period from three independent data sets and with objective criteria:  First, all the ground-based data gives the highest DFT peak at 1.24$\pm$0.02 days.  Second, the $K2$ data set has its highest DFT peak (twice as high as any other peak) at 1.24 days, even though this peak arises from the Superflare flux.  Third, the SED from GALEX to WISE (0.231--4.6 microns) has a cutoff in the near infrared that records the outer-edge of the accretion disk, and this forces $P$ to be from 1.1--2.5 days, with the lower edge of the range simply for the accretion disk nearly filling its Roche lobe.  Taken alone, each of the first two data sets is not convincing, but the coincidence between all three is convincing.  So we conclude that $P$ is 1.24$\pm$0.02 days for V2487 Oph.

\subsection{White Dwarf Mass}

The white dwarf must be more massive than something like 1.20 M$_{\odot}$, as otherwise the recurrence time-scale cannot be shorter than 98 years.  For the preferred recurrence time-scale of 18 years (Schaefer 2010), the white dwarf mass must be $M_{WD}$$>$1. 25 M$_{\odot}$ (Shen \& Bildsten 2009).  Shara et al. (2018) used the observed outburst amplitude and the decline rate ($t_2$) to derive that $M_{WD}$=1.35 M$_{\odot}$.  Hachisu et al. (2002) modeled the light curve with a derived $M_{WD}$=1.35$\pm$0.10 M$_{\odot}$.

\subsection{Distance}

Historically, the distance to V2487 Oph could only be guessed by the now-discredited maximum-magnitude-rate-of-decline (MMRD) relation.  But for our RN, the claimed distance was ridiculously outside our Milky Way (Schaefer 2010), with this being one of many reason to know that the MMRD can never be trustworthy.  (Once {\it Gaia} distances became known for many novae, the MMRD was demonstrated to be incorrect even to an order-of-magnitude, see Schaefer 2018.)  Distance methods using extinction measures are not useful for V2487 Oph because the nova is far past all galactic dust so only a useless lower limit is possible anyway.  And for the distance method involving measured parallaxes, the first two data releases for the {\it Gaia} satellite had very large error bars for V2487 Oph, for these being useless.  This sets up the problem where published distances have spanned the range 1120 to 32400 parsecs.

Another method to investigate is that of deriving a blackbody distance to the companion star.  Going into our study, we had expected to pull out an orbital period (and hence get a good estimate of the size of the companion star) and expected to pull out a temperature and flux from the companion star from the SED.  With these, we could calculate an accurate distance, as we did for RNe U Sco, T CrB, RS Oph, V745 Sco, and V3890 Sgr (Schaefer 2010).  Alas, our SED analysis only places poor limits on the companion's flux and temperature.  So this hopeful method has failed completely.

Now, we have the improved {\it Gaia} EDR3 data release{\footnote{https://archives.esac.esa.int/gaia}}.  The latest parallax for V2487 Oph is 0.129$\pm$0.075 milli-arc-seconds.  To translate this parallax into a distance, we must use the prescribed `exponentially decreasing space density' (EDSD) priors (Luri et al. 2018).  Unfortunately, in this case with large relative error bars for the parallax, the one-sigma range changes substantially with the expected distance scale ($L$).  V2487 Oph is from a bulge population, then $L$=8000 pc is appropriate, and we then derive a distance of 11000$_{-1600}^{+20600}$ parsecs.  Given the size of our Milky Way and the very large error bars, the {\it Gaia} EDR3 parallax is effectively only giving a lower limit on the distance.  With the EDSD prior, the most probable distance is 11,000 pc, with the middle 68 per cent probability all with distance greater than 9400 pc.  If we push to a higher confidence level, the two-sigma lower limit is 7800 pc.  So in all, to good confidence, with the {\t Gaia} EDR3 parallax, we can only say that V2487 Oph has a distance $\gtrsim$7800 pc.

We can use the galactic position of V2487 Oph to get a convincing range of distances.  V2487 Oph has a galactic longitude of 6.60$\degr$ and a latitude of +7.78$\degr$, which is close to the galactic centre.  With the {\it Gaia} distance limit putting our RN far outside the galactic plane, we know that V2487 Oph is from the bulge population.  The bulge population of novae is entirely contained within a radius of 15$\degr$ from the galactic centre (e.g., Warner 2008), so the maximum distance from the galactic centre is 2000 pc.  So we strongly expect V2487 Oph to have a distance of 8000$\pm$2000 pc in round numbers.

So where does this leave us for the distance?  Well, many of the old and new methods have failed to produce useable constraints.  The only useful and reliable information is that the {\it Gaia} EDR3 parallax gives 11000$_{-1600}^{+20600}$ parsecs, and the galactic position puts V2487 Oph into a bulge population and bulge novae all have distances of 8000$\pm$2000 pc.  In round numbers, we conclude that the distance to V2487 Oph is best approximated to be 8000$\pm$2000 pc.

\subsection{Extinction}

Based on their observations of the O {\rm I} lines at 8446 and 11287 \AA, Lynch et al. (2000) found $E(B-V)$=0.38$\pm$0.08.  Based on an adopted $B-V$=+0.23 at peak, then $E(B-V)$=0.4 (Schaefer 2010).  Based on an adopted $B-V$=$-$0.02 when the light curve has faded 2 mag below peak, then $E(B-V)$=0.6 (Schaefer 2010).  For the galactic position of V2487 Oph, the entire line-of-sight through our galaxy has $E(B-V)$=0.6529$\pm$0.0424 (Schlafly \& Finkbeiner 2011), so the nova itself must have $E(B-V)$$\leq$0.65.  Given the $>$7800 pc distance (see previous Section) putting the nova far outside of any galactic dust, we can only take the maximum extinction value.  So our final estimate of the extinction to V2487 Oph is $E(B-V)$=0.65$\pm$0.04.  This translates, with $R$=3.1 (as appropriate for our nearby galaxy), to a light loss in the V-band of $A_V$=2.01$\pm$0.12.

\subsection{Absolute Magnitude}

The absolute magnitudes in the V-band, in peak and in quiescence, are useful properties to know.  For this, we can do no better than to take the tabulated apparent magnitudes, in peak and quiescence, from Schaefer (2010).  We also need an extinction, for which we will use the reliable $A_V$=2.01$\pm$0.12 from the previous Section.  We also need a distance, for which we will use the 8000$\pm$2000 pc estimate, as based on the {\it Gaia} parallax and the result that V2487 Oph is in the bulge population of novae.

The peak magnitude of V2487 Oph is $V_{peak}$=9.5, with an uncertainty of under 0.1 mag.  Then, we have the absolute magnitude at peak of $M_{V,peak}$ equal to $-$7.0.  The uncertainty is dominated by the distance error bars, and is $\pm$0.6 mag.  This value is right at the centre of the overall distribution for novae in general (Schaefer 2018), so there is nothing surprising here.

The magnitude in quiescence is $V_q$=17.3, with a range of 0.2 mag.  Then the absolute magnitude  in quiescence is $M_{V,q}$ is $+$0.8$\pm$0.6.  This is greatly more luminous than almost all CVs.  Patterson et al. (2021) point out that all classical novae and dwarf novae are less luminous than $M_{V,q}$=$+$3.0, while recurrent novae were all spread from $+$3.0 to 0.0.  So V2487 Oph fits nicely into its RN class.

\subsection{Accretion Rate}

All RNe must have high accretion, as that is the only way to pile mass on to the white dwarf fast enough to have a recurrence time-scale of under one-century.  For any RN, the accretion rate must be 2--20 $\times$10$^{-8}$ M$_{\odot}$ yr$^{-1}$ while an 18 year cycle means 8--20 $\times$10$^{-8}$ M$_{\odot}$ yr$^{-1}$ (Shen \& Bildsten 2009).  Further, the modeling of V2487 Oph itself by Shara et al. (2018) gives the accretion rate to be $\dot{M}$=4$\times$10$^{-8}$ M$_{\odot}$ yr$^{-1}$.  The V2487 Oph model of Hachisu et al. (2002) returns $\dot{M}$$\sim$15$\times$10$^{-8}$ M$_{\odot}$ yr$^{-1}$.

The {\it shape} of the SED gives a reasonable value for $\dot{M}$.  From Section 2.3, we see that the companion star provides little flux, so the shape of the SED is dominated by the disc.  A difficulty of getting the accretion rate is that the outer radius of the accretion disc is not independently known.  For the calculated disc-only model in Fig. 5, we assumed that the disc filled most of the white dwarf's Roche lobe for a 1.24 day orbit.  With this, the best-fitting has $\dot{M}$=5$\times$10$^{-8}$ M$_{\odot}$ yr$^{-1}$.  The uncertainties on this are hard to quantify (due to possible systematic errors in the data and due to having only poor constraints on the disc size), but we see little chance that the accretion rate is outside the range of (1--9)$\times$10$^{-8}$ M$_{\odot}$ yr$^{-1}$.

The best measure of the accretion rate comes from the absolute magnitude of the disc.  In the V-band, almost all of the light is from the disc, so we have $M_{V,disc}$=$+$0.8$\pm$0.6.  Within the alpha-disc model, we can use this {\it luminosity} (not the SED {\it shape}) to estimate the accretion rate.  The conversion from absolute magnitude to $\dot{M}$ has additional uncertainties, primarily from not knowing the inclination angle of the disc.  The lack of photometric modulations on the orbital period suggests that the inclination is small, so we will consider the range 0$\degr$ to 30$\degr$.  With these ranges of inclination and absolute magnitude, our alpha-disc model returns a range of (8--57)$\times$10$^{-8}$ M$_{\odot}$ yr$^{-1}$, and a middle value of 20$\times$10$^{-8}$ M$_{\odot}$ yr$^{-1}$.

So we have many ways to estimate $\dot{M}$, and they are all consistent to within their error bars.  The centre of all the overlap is $\dot{M}$=9$\times$10$^{-8}$ M$_{\odot}$ yr$^{-1}$, which is our best estimate.  However, there is only poor accuracy, so we would take the realistic range of accretion rates to be (3--16)$\times$10$^{-8}$ M$_{\odot}$ yr$^{-1}$.  Again, this is a very high accretion rate for any CV, even getting near the 'steady hydrogen burning' regime (Nomoto 1982; Shen \& Bildsten 2009), with this being the hallmark of RNe.  This very-high $\dot{M}$ is a requirement for all RNe, as only by spilling gas on to the white dwarf at immense rates can the trigger mass be accumulated in under 100 years.

\section{Superflare Relations}

The basic properties of the V2487 Oph Superflares are given in Table 5.  In an effort to get clues as to the mechanism, we have sought correlations and distributions amongst the tabulated properties.  We have three results that are relevant for distinguishing between possible mechanisms and scenarios.

\subsection{Superflare Energies Are Proportional to the Time to the Next Flare}

The time from one flare to the next, $\Delta T$, has an approximately exponential distribution, which is consistent with the flare times being random.  Within this constraint, it is still possible for $\Delta T$ to depend on some flare property, with that property being random.  With the precedents of Type {\rm I} and Type {\rm II} X-ray bursts, the waiting time from one burst to the next is proportional to the burst energy of the first burst.  (X-ray bursts are a possible analog for fast flares produced by gas in an interacting binary falling onto a compact star, a neutron star in this case.  The common Type {\rm I} bursts are powered by runaway thermonuclear burning of hydrogen-rich gas accumulated on the surface of the neutron star, while Type {\rm II} bursts are powered by the sudden release of gravitational energy as gas bundles drops onto the neutron star.  Both of these types of X-ray bursts have the flare energy scaling as the mass depleted from a reservoir, while the time until the next flare is the time it takes to refill the reservoir up to some trigger threshold.  Historically, these arguments were critical for understanding X-ray bursts.)  So we should check whether the V2487 Oph Superflares have any significant relation between $E$ and $\Delta T$.

For V2487 Oph, the logarithms of $E$ and $\Delta T$ from Table 5 are plotted in Figure 6.  We see a strong correlation, where the more energetic the flare, the longer the wait time until the next flare.  This correlation is formally significant (with correlation coefficient of +0.70 for 58 points), but it still has a lot of scatter.  The measurement uncertainties are greatly smaller than the intrinsic scatter, and we cannot apportion the error bars between the two parameters.  The amount of scatter causes a large and poorly-defined uncertainty in the slope.  Near the centre of this range is the slope of unity (the gray line in Figure 6).  On a plot of two logarithmic values, the slope of unity is special, as it means that the two quantities are in a proportionality relation.  That is, $E$$\propto$$\Delta T$.  So the V2487 Oph Superflares have the loose property that a low-energy event is soon followed by another flare, while a high-energy event makes for a long delay until the next flare.

The substantial scatter seen in Fig. 6 implies that the $\Delta T$ values are determined by more parameters than just $E$.  But the strong correlation provides evidence as to the nature of the flare mechanism, and it provides a constraint that any viable mechanism must explain.

\begin{figure}
	\includegraphics[width=1.05\columnwidth]{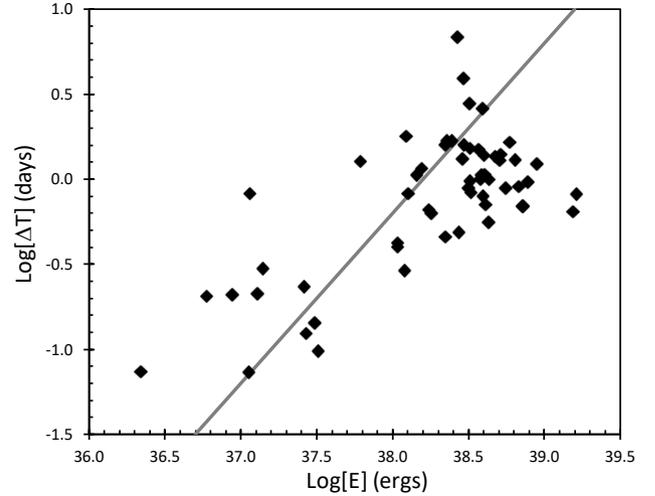}
    \caption{The Superflare energy versus the time between flares.  The flare energy, $\log [E]$, is positively correlated with the time interval to the next flare, $\log [\Delta T]$.  This correlation is highly significant, but there is substantial scatter.  The large scatter leads to a large uncertainty in the best-fitting slope.  The gray line shows a slope of unity, which is close to the middle of the range of uncertainty.  A slope of unity means that the energy is proportional to the wait time.  That is, a relatively energetic flare has a long wait time until the next flare, and a relatively weak flare is swiftly followed by the next flare.  This behavior is characteristic of some reservoir of energy being steadily filled, until it reaches some critical threshold, then dumps some part of that energy, only to have the refilling of the reservoir continue until the next flare.}  
\end{figure}

\subsection{Superflare Durations are a Power Law of Flare Energies}

Observations of many stellar systems with magnetic reconnection flares show that the flare durations are a power law of the flare energies, with an apparently universal relation that the duration is proportional to something like $E^{+0.42}$.  We checked to see how the durations are related to flare energy for V2487 Oph Superflares.  Figure 7 shows the plot of the data from Table 5, with logarithms of $D_{eq}$ versus $E$.  We see a strong and highly significant correlation, with only modest scatter.  The best-fitting slope is $+0.44\pm0.03$.  This strong result should be compared against predictions from all the various models and scenarios.

\begin{figure}
	\includegraphics[width=1.05\columnwidth]{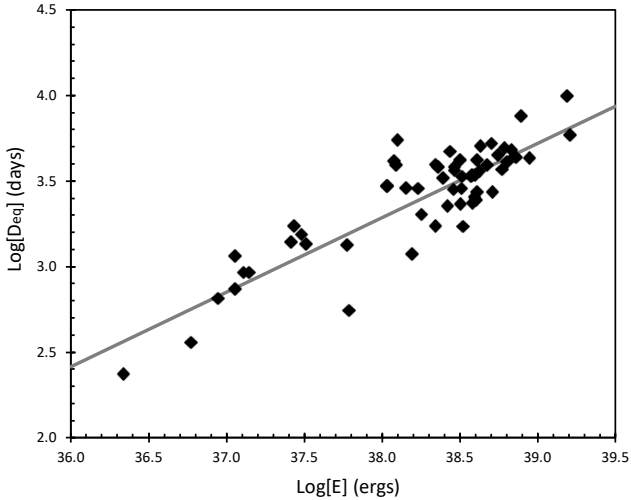}
    \caption{Superflare durations scale close to $E^{+0.42}$.  This logarithmic plot of $D_{eq}$ versus $E$ shows a strong correlation between the duration and the Superflare energy, $D_{eq}$$\propto$$E^{0.44\pm0.03}$, displayed as a gray line.  Critically, this is closely similar to the many cases where magnetic reconnections on stars make for a wide-ranging power law of similar indices.  White light solar flares have an index of 0.38$\pm$0.06, ordinary flare stars have indices of 0.47$\pm$0.07, Superflare stars have 0.39$\pm$0.03 with {\it Kepler} data and 0.42$\pm$0.01 with {\it TESS} data.  So it appears that magnetic reconnection on stars produces something like a universal law with $D$$\propto$$E^{+0.42}$.  This relation is not seen or expected from flares with a gravitational energy source or with a nuclear burning source.  So the fact that V2487 Oph Superflares follow this distinct universal relation is strong evidence that they are powered by magnetic reconnection.}  
\end{figure}

 \subsection{Distribution of Superflare Energies}

The distribution of Superflare energy provides another relation that can be compared to scenarios and models.  This distribution is usually quantified as $\frac{dN}{dE}$, the number of Superflares with $E$ in some range of width $dE$, plotted on a log-log plot, and fitted to a power law of $E^{-\alpha}$.  The error bars are simply from propagating the Poisson uncertainties on the counts in each bin.  Always, the $\frac{dN}{dE}$ starts deviating from the power law to the left of some completeness threshold, while the right side of the plot starts having large vertical error bars due to the scarcity of the most-energetic bursts.  For V2487 Oph, the plot turns over to low-$E$ for $\log[E]$$<$38.4, while the highest $\log[E]$ is 39.21.  This leaves only a relatively small dynamic range over which to measure $\alpha$.  With only 34 Superflares in this power law regime, the uncertainty on $\alpha$ will be relatively large.  WIth this, the best-fitting value for $\alpha$ is 2.34$\pm$0.35.

\section{The Nature of the Superflares}

The nature of the Superflares is now the central mystery of V2487 Oph.  With no Superflares having ever been seen on any CV (or any contact binary of any type), we have no close precedents to guide us to a reasonable model.  Still, there are only three energy sources that have any real chance of powering the Superflares; the gravitational energy of accreting material, nuclear energy from thermonuclear burning of material on the surface of the white dwarf, and magnetic energy from the reconnection of field lines in the volume between the white dwarf and the companion.  In this Section, we will consider the basic physics of these three energy sources, trying to pick out what is possible and what is reasonable.

Let us start by giving details on the expected scenario for each of the three possible energy sources.  We expect that upcoming work by specialists can substantially sharpen the scenarios.

The first possible energy source is gravitational energy of material falling off the companion star, through the accretion disc, and on to the white dwarf.  Our scenario for this is that inflowing gas is  suddenly dropped to a lower orbit, the gravitational potential gets converted to thermal energy in the gas, and this then radiates the Superflare light.  Just such a physical mechanism is well known as the cause of dwarf nova eruptions, where the material in the accretion disc moves inward slowly until a critical threshold of disc temperature is reached, whereupon the gas viscosity increases so that the gas infalls at a fast rate, suddenly freeing up a lot of gravitational energy.  A second known example is the flux of the `hotspot' (where the accretion stream leaving the companion star hits the outer edge of the disc), with the gravitation energy of the falling gas gets converted to thermal energy, then to radiative energy.  In the gravitational-energy scenario, we could imagine some instability in the inner disc that suddenly dumps gas on to the white dwarf, but such an idea has the issue that no such instability is known to work around a white dwarf (Lewin et al. 1993; Frank et al. 2002).

The second possible energy source is the nuclear energy from thermonuclear burning of gas accreted on to the the surface of the white dwarf.  Nuclear burning is very efficient, as ordinary hydrogen fuel converts 0.7 per cent of its mass into energy.  Once triggered, the initial burning will heat up nearby fuel past its kindling temperature, and the resultant runaway can power some sort of a flare, lasting until the fuel is exhausted.  A nova eruption is exactly this scenario.  Something drastic must change from a nova to a Superflare, so as to get the greatly shorter duration and greatly smaller energy.  Our only plausible idea is that perhaps some fraction of the accreted material is funneled on to a very small area of the white dwarf surface (say, by complex magnetic field lines), which builds up a layer that reaches trigger conditions at its base, and then it undergoes a normal thermonuclear runaway over that small area alone.  Material would keep accreting into this focused area, making for repeated eruptions.

The third possible energy source is reconnection of magnetic field lines in the volume between the white dwarf and the companion star.  The physical scenario is that the field lines get twisted (by the bases of the lines being trapped in either a rotating orbit or in the accretion disc), the field lines undergo magnetic reconnection when the stresses reach a critical breaking point, the reorganized field lines will then create very energetic Alfven waves that accelerate ambient electrons to very high energies, these very energetic electrons slam into their base (tied to the white dwarf, the accretion disc, or the companion star), the electrons release a sudden large impulse of energy by the bremsstrahlung processes, and this liberated radiation heats the gas which produces the optical light of the Superflare.  The magnetic fields keep getting amplified by the rotation of the disc/companion, and the reconnections are triggered when the magnetic stress reaches the breaking threshold.  

Most of the physics of this basic process is well known in a variety of settings, and it is essentially identical across the many types of stellar magnetic-reconnection events.  Naturally, the process is best studied for solar flares on our own Sun, as we can watch in real time the resolved structure change and develop with light of wavelengths from radio to gamma-ray (Haisch, Strong, \& Rodono 1991).  In this case, the magnetic loops have both their footprints firmly attached to the roiling gas around a sunspot group, with this motion causing the field lines to be stretched and twisted (Janvier 2017).  Hesse \& Cassak (2020) give a general review on reconnection is astrophysical settings.  The theory of reconnection is still poorly known on the smallest size-scale for the time when field lines change their connectivity, and this lack of understanding of the microphysics is beyond current numerical analysis and beyond current laboratory testing.  This lack of understanding means that no one has any accurate threshold for when reconnection is triggered.  Nevertheless, reconnection does occur, it occurs in many stellar astrophysics settings, and its consequences are well understood physics.  The same overall mechanism as for solar flares is thought to apply to ordinary flare stars and to isolated superflare stars (Shibata et al. 2013).  This flare mechanism does not require that the footprints of the loops in the field lines be anchored to footprints in the same star.  So the model for Superflares on RS CVn stars is that the magnetic field lines are anchored to the two stars in the binary, with these lines then getting stretched and twisted by the orbital motion (Simon et al. 1980; Gunn et al. 1997).  For RS CVn stars, the B-field originates in the G-type companion star, where the characteristically high stellar activity and magnetic field are due to the star being spun-up, as forced by co-rotation with the orbit (Rodono 1992).  This is the same situation as for V2487 Oph, where the G-type companion star is spun-up to a rotation period of 1.24 days, with this very fast rate necessarily making for high activity, large starspots, and intense magnetic fields,  So, just as there are superflares on RS CVn stars, we could expect that there can be superflares on V2487 Oph.  A difference is that the RN has gas flowing off the star across the inner Lagrangian point.  This ordinary CV mass transfer would carry along with it any field lines that were anchored to it before falling through the Roche lobe, and this would provide a seed magnetic field that will be amplified by the gas as it rotates around in the accretion disc.  The outcome will be magnetic field lines with its footprints in the companion star and the accretion disk, with the inevitable twisting and stretching as the gas and star rotate madly nearly once a day.  Then, while we do not know the physics at the point and time of the reconnection, the field lines must break, Alfven waves must drive a relativistic jet of ambient electrons that will smash into the footprints, releasing the Superflare energy.

For the magnetic reconnection energy source scenario, there are abundant precedents for isolated-in-time, roughly-daily, highly-energetic, roughly-hour-long flares with fast-rises and `exponential' tails, all in stellar settings with energies up to 10$^{38}$ ergs.   For our own Sun, white-light solar flares (WLSFs) have been known since the Carrington Event of 1859, and these involve magnetic reconnection in large loops grounded in sunspots.  For low-mass main sequence stars, spectral classes K and M, flare stars are common, also powered by magnetic reconnection.  Superflares (as greatly more energetic than normal flares) also occur on ordinary very-low mass main sequence stars, for example on Proxima Centauri (Howard et al. 2018), L-type dwarfs (Paudel et al. 2020), plus two late-M stars with Superflare amplitudes 9.5--16 mag (Schaefer 1990).  For G-type main sequence stars, Superflares are seen with energies up to 2$\times$10$^{38}$ ergs in field stars (Schaefer et al. 2000), and are very-well observed on many solar-like stars inside the {\it Kelper} field (Maehara et al. 2012).  Eerily, most of these Superflare stars are single, not-young, and indistinguishable from our own Sun.  These Superflares are now well-established to be due to the reconnection of loops of magnetic fields, even though the field configuration and location of the footprints are not known (Rubenstein \& Schaefer 2000; Maehara et al. 2012; Shibayama et al. 2013; Namekata et al. 2017).  Superflares are also common on A-type and F-type main sequence stars (Balona 2012), on Mira stars (Schaefer 1991; de Laverny et al. 1998), and a wide variety of massive main sequence stars, giants, and supergiants (Schaefer 1989).  The ubiquity of Superflares in a very wide variety of stellar situations proves that the mechanism is very robust, so Superflares can easily be expected from most any type of star.  For stars like the components in the V2487 Oph binary, white dwarfs are known to have superflares (Schaefer 1989), and a substantial fraction of G-type sub-giants are known to have superflares (Yang \& Liu 2019).  With white dwarfs and G-type sub-giants both being susceptible to Superflares, we should not be surprised if V2487 Oph has Superflares.  Nevertheless, there is a stunning absence of Superflares on any type of interacting binary, including novae, RNe, CVs of all types, and neutron star binaries of all types.  This is despite a huge investment in optical and X-ray photometry of a myriad of such stars for many decades.  This makes V2487 Oph into a startling and special case, as the only known Superflare star that is an interacting binary.

\begin{table*}
	\centering
	\caption{Testing Energy-Source Models Against Observations}
	\begin{tabular}{lllll} 
		\hline
		Property & V2487 Oph & Gravitational Energy Scenario  & Nuclear Energy Scenario  &  Magnetic Energy Scenario   \\
		\hline
Energetics	&	$\gtrsim$10$^{38}$ ergs/flare	&	\checkmark ~ $VB^2/8\pi$$<$10$^{48}$ erg	&	\checkmark ~ 0.0007$\dot{M}$ $\Rightarrow$ 10$^{41}$ ergs/year	&	\checkmark ~ 10$^{-11}$ M$_{\odot}$/yr  on 400-m area	\\
	&	$\sim$10$^{41}$ ergs/year	&		&		&	~~~~~~~~$\Rightarrow$ 10$^{41}$ ergs/year	\\
$E$ distribution	&	$\frac{dN}{dE}$$\propto$$E^{-2.34 \pm 0.35}$	&	$\times$ ~ DN:~monoenergetic or bimodal	&	$\times$ ~ Novae (RN):~monoenergetic	&	\checkmark ~ Universal law: ~ $E^{-2.0\pm0.2}$	\\
	&		&	$\times$ ~ Type {\rm II} XRB:~~$E^{-1.03}$	&	$\times$ ~ Type {\rm I} XRB:~ $\frac{dN}{dE}$ constant	&		\\
	&		&		&	~~~~~~~~over small range	&		\\
$\Delta T$ between 	&	$\Delta T$$\propto$$E$	&	$\times$ ~ DN:~no correlation	&	$\times$ ~ Novae (RNe):~no correlation	&	? ~ $\Delta T$$\propto$$E$ expected if only	\\
~~~~flares	&		&	? ~ Type {\rm II} XRB:~$\Delta T$$\propto$$E$	&	\checkmark ~ Type {\rm I} XRB:~ $\Delta T$$\propto$$E$	&	~~~~~~~~{\it one} reservoir is active,	\\
	&		&		&		&	~~~~~~~~like smearing in the disk	\\
Rise times	&	$\tau_{rise}$$\lesssim$1 min.	&	$\times$ ~ DN:~$\tau_{rise}$$\sim$1 day	&	\checkmark ~ Inner disk:~$\tau_{rise}$$\sim$3 sec.	&	\checkmark ~ WLSF: ~ 0.1--10 min.	\\
	&		&	\checkmark ~ Type {\rm II} XRB:~$\tau_{rise}$$\sim$1 sec.	&		&	\checkmark ~ Flare Stars: ~ 0.2--5 min.	\\
	&		&		&		&	\checkmark ~ Superflare Stars: ~ 1--10 min.	\\
Light curve	&	Spike $+$ tail	&	$\times$ ~ DN:~rounded/symettrical LCs	&	\checkmark ~ Novae:~ Spike$+$tail	&	\checkmark ~ WLSF: ~ Spike$+$tail	\\
~~~~shapes	&		&	~~~~~~~~with no spikes or tails	&	\checkmark ~ Type {\rm I} XRB:~  Spike$+$tail	&	\checkmark ~ Flare Stars: ~ Spike$+$tail	\\
	&		&	$\times$ ~ Type {\rm II} XRB:~rounded LCs, 	&	\checkmark ~ Burning on WD:~ Spike$+$tail	&	\checkmark ~ Superflare Stars: ~ Spike$+$tail	\\
	&		&	~~~~~~~~with no spikes or tails	&		&		\\
Durations	&	$D_{eq}$$\propto$$E^{0.44\pm0.03}$	&	$\times$ ~ DN:~$D$$\propto$$E^{0.598\pm0.004}$	&	$\times$ ~ Novae (RN):~$D$ constant	&	\checkmark ~ WLSF: ~ $D$$\propto$$E^{0.38\pm0.06}$	\\
	&		&	$\times$ ~ Type {\rm II} XRB:~~$D$$\propto$$E^{1.3\pm0.3}$	&	$\times$ ~ Type {\rm I} XRB:~ $D$ constant	&	\checkmark ~ Flare Stars: ~ $D$$\propto$$E^{0.47\pm0.07}$	\\
	&		&		&	$\times$ ~ Burning on WD:~$D$$\propto$$E^{0.0}$	&	\checkmark ~ Superflare Stars: ~ $D$$\propto$$E^{0.39\pm0.03}$	\\
	&		&		&		&	\checkmark ~ Superflare Stars: ~ $D$$\propto$$E^{0.42\pm0.01}$	\\
		\hline
	\end{tabular}
\end{table*}

\section{Testing the Energy-Source Models}

We have only three possible energy sources, and these should be tested against the properties of the V2487 Oph Superflares.  Unfortunately, the various scenarios usually do not have a ready way to make predictions for specific properties from first principles.  The only way to make many of these tests is to look at the closest analogue systems.  In this section, we look at the predicted Superflare properties (based on either first principle calculations or analogue systems), then compare these predictions versus the actual observations from V2487 Oph.  

There are many detailed comparisons and analyses in this Section, so we have collected them together into Table 7.  For each predicted item for each energy-scenario, we have characterized the agreement with V2487 Oph either as ``\checkmark" for good agreement, ``?" for unclear cases, and ``$\times$" for disagreement.  Readers can rapidly scan down the columns for each energy source and see the successes and failures to predict the properties of V2487 Oph.

\subsection{Superflare Energetics}

The energetics of the Superflares will set the stage for what is possible or impossible in the three scenarios.  What we need to explain is a single flare energy of over 10$^{38}$ ergs, while the yearly energy requirement is of order 10$^{41}$ ergs.

The energy available for the gravitational energy scenario works off the gravitational potential energy of a mass $M$ being released as it moves inward towards the white dwarf.  The potential energy is the usual $-G M_{WD} M / R$, with the mass at radius $R$.  The released energy will depend on the initial and final values of $R$.  If the entire potential energy of 6$\times$10$^{-14}$ M$_{\odot}$ is released then we can power a typical Superflare at 10$^{38}$ ergs.  As such, an accretion rate of 6$\times$10$^{-11}$ M$_{\odot}$ yr$^{-1}$ will power the Superflares.  This is $\sim$0.0007 of the total accretion in V2487 Oph.  But our various scenarios will have only a fraction of the potential energy released.  The fraction released is very low when we are looking at the energy of a blob of gas hitting the outer edge of the accretion disc, so low that it would require more than the entire accretion rate to power just the Superflares.  So this energetics limit shows that Superflares are not from gas packets in the accretion stream hitting the hotspot at the edge of the disc.  However, for the case of gas in the inner regions of the disc falling on to the white dwarf, roughly half of the original potential energy will be available.  In this case, only $\sim$0.1 per cent of the entire accretion rate of V2487 Oph is needed to power the Superflares. 

The energy available for thermonuclear burning of a mass M$_{burn}$ of hydrogen is $0.007M_{burn}c^2$.  A single Superflare needs a burning mass of order 10$^{-14}$ M$_{\odot}$ for fuel, while a whole year needs of order 10$^{-11}$ M$_{\odot}$ for fuel.  This is small compared to the accretion of near 10$^{-7}$ M$_{\odot}$ yr$^{-1}$.  So only $\sim$0.0001 of the infalling mass needs to be burnt for Superflares.  But how can such a small mass be heated to trigger the runaway?  We can only think that the 10$^{-14}$ M$_{\odot}$ of fuel gets piled on to a small area, so that in a day it becomes deep enough  to reach the trigger condition at the base of the fresh material.  How small must this area be?  Well, the radius of the 1.35 M$_{\odot}$ white dwarf is near 0.003 R$_{\odot}$ and the ignition mass is 10$^{-5.8}$ M$_{\odot}$ spread over the entire surface (Shen \& Bildsten 2009).  If the hypothetical small region occupies only $\sim$10$^{-8}$ of the white dwarf surface area to the same depth, then the trigger condition will be met with enough available energy to power a single Superflare.  This is an area with a radius of around 400 m.  So the energetics and the runaway trigger conditions are satisfied if just 10$^{-11}$ M$_{\odot}$ yr$^{-1}$ is channeled into this 400 m radius area.

The energy available for the magnetic reconnection scenario depends on the magnetic field, $B$, and the volume of the field, $V$, that is annihilated.  The available magnetic energy is $(B^2/8\pi)V$.  Unfortunately, we have only very loose constraints on $B$ and $V$.  The magnetic fields on some solar-like Superflare stars are measured to be over 1000 G (Schaefer et al. 2000), while the companion star in V2487 Oph is spun up to a rotation period of 1.24 days and could create relatively high surface fields and stellar activity.  The white dwarf in V2487 Oph could well have a substantial magnetic field, although this field cannot be as high as for polar CVs, because we see much of the accretion disk in the SED, and we see flickering.  Still, V2487 Oph might be an intermediate polar with a magnetically truncated inner disk, with a surface magnetic field of up to roughly 10$^7$ G.  In addition, fields grounded on the companion star or the accretion disc will be endlessly amplified by the stress induced from the rotations.  So $B$ values up to 10$^7$ G are plausible, and possibly much higher with the expected amplification.  The volume will be some loop length times some cross section of the loop that is annihilated.  The loop length has an upper limit near the radius of the white dwarf Roche lobe.  For V2487 Oph with expected stellar masses and $P$=1.24 days, the maximal loop length is 3$\times$10$^{11}$ cm.  We have no useable idea or experience as to the cross section of the loop to be annihilated, and this will change substantially from the `footprint' to the middle of the loop.  If the white dwarf is grounding one end of the loop, then the footprint size can only be less than some modest fraction of the white dwarf radius of 2$\times$10$^{8}$ cm.  If the loop is grounded to the companion star, perhaps a 6000 K sub-giant star spun up a rotation period of 1.24 days, then the footprint size might be capped at a size similar to a large sunspot on our own Sun, so a maximum cross sectional radius might be 10$^{10}$ cm.  If the loop is grounded in the accretion disc, even a small initial cross section would expand rapidly from the shear in the Keplerian velocities across the footprint.  Let us now calculate the available magnetic energy for two cases, first the maximal energy, and then what might be a typical energy.  For the maximal case, pushing everything to its upper limits ($B$=10$^9$ G, loop length of 3$\times$10$^{11}$ cm, loop cross section of 10$^{20}$ cm$^2$), we have 10$^{48}$ ergs of available magnetic energy.  This maximal case might not be reachable in practice.  We expect that the magnetic field cannot be maintained as high as a billion Gauss over a large volume.  (The field lines with footprints in the disk or on the companion will necessarily be speedily twisted and stretched until they reach some poorly-determined threshold for magnetic reconnection.)  But this is the whole point of the scenario, that the magnetic fields are increased by rotation until some threshold is reached, whereupon the magnetic fields reconnect to a lower energy configuration, and that is what releases the energy to make the Superflare.  Still, this maximal case demonstrates that the magnetic reconnection scenario can have many orders-of-magnitude more energy available than needed for Superflares.  For a more modest perhaps-typical case (with a million Gauss field, a loop 10$^{11}$ cm long with a cross sectional size of 10$^9$ cm), the energy from reconnection would be 4$\times$10$^{39}$ erg, comparable to an energetic Superflare.  The point is that the magnetic reconnection scenario has sufficient available energy to power the V2487 Oph case.

In summary, all three possible energy sources (gravitational, nuclear, and magnetic) can have many orders-of-magnitude more available energy than is needed for the V2487 Oph superlfares.

\subsection{Distribution of Superflare Energies}

We have measured that the V2487 Oph Superflare energies follow the distribution ($dN$/$dE$)$\propto$$E^{-2.34 \pm 0.35}$.  This power-law exponent can be compared to those observed or expected for each of the three energy sources.

For the gravitational energy scenario, we know of no general law or calculation that fits all circumstances.  So the best we can do is to look at the power-law exponents for situations with accretion instabilities, in particular dwarf nova events and Type {\rm II} X-ray bursts.  For dwarf nova (DN) events, individual outbursts have total energies that are uni-modal or bi-modal in distribution.  The bi-modal cases are when a single CV has so-called `short' and `long' eruption, and when the dwarf nova has both regular outbursts and `super-outbursts'.  Within each mode the energies cluster around some constant (different for each CV), with roughly a factor-of-two scatter.  To provide a specific confirmation, we have analyzed 704 outbursts from the prototype dwarf nova SS Cyg, and we find that essentially all eruptions (including both short and long) cover an energy range of a factor-of-six with a flat $dN/dE$ that cuts off completely and abruptly (over a range of 10 per cent of the energy).  The point is that the dwarf nova accretion instability events have a physical cutoff in energy, with all this being far from the roughly -2 power law seen for V2487 Oph.  For Type {\rm II} X-ray bursts (XRBs), the energy distribution of the $E$ values in Kunieda et al. (1984) has a power law index of -1.03$\pm$0.10 up until a cutoff $E$.  This sharp cutoff is highly significant, with the top bin with width 0.2 in $\log[E]$ having 16 bursts, while there are zero bursts with any higher energy.  This energy distribution is certainly greatly different from that seen for V2487 Oph.  

For the nuclear energy scenario, we can only look to the closest physical systems to the hypothetical thermonuclear runaway idea.  For nova eruptions, RNe are the only case where we can look at distributions of energy for events with similar conditions.  For RN eruptions from all individual systems, we have the strong result that the light curves are all effectively identical (Schaefer 2010).  So the distribution of $E$ is mono-energetic.  For Type {\rm II} X-ray bursts, the distributions of $E$ for each burst source are roughly a flat-top distribution with cut offs to high-$E$ and low-$E$ (Lewin et al. 1993).  That is, burst energies fall equally throughout a fairly narrow range.  The size of this range varies with the burster, but it ranges from 2$\times$ to 5$\times$.  The two closest analogs for nuclear powered flares both fail to produce a power law distribution of flare energies as seen for V2487 Oph.

For the magnetic reconnection scenario, Shibayama et al. (2013) and Maehara et al. (2015) note that a power law index from -1.8 to -2.2 is a universal property of stellar reconnection flares.  That is, for over 12 orders-of-magnitude in stellar flare energy (from 10$^{24}$ to 10$^{36}$ ergs), the frequency of flares follows the same power law distribution.  Indeed, they are using this as additional evidence that Superflares are magnetic reconnection events.  With this universality, the power law shape and power law index for the $dN/dE$ distribution became a prediction to be tested for stellar magnetic reconnection models.  Taking all the many types of stellar reconnection events (from solar nano flares up to the bigger Superflares on normal stars), the frequency distributions lie along a band with a power law index of -1.8.   For normal M-type flare stars, we find an average power law index of -1.88 with the flare data in Yang \& Liu (2019).  For Superflares on G-type main sequence stars, Shibayama et al. (2013) found a power law index of -2.2. The magnetic reconnection scenario makes a prediction that the power law index will be roughly -1.8 to -2.2, and this is easily consistent with what is seen for the V2487 Oph Superflares.

In summary, scenarios involving the gravitational and nuclear energy sources empirically do not produce anything like the observed energy distribution for V2487 Oph, while the magnetic reconnection scenario has the successful prediction for the $dN/dE$ power law index that matches that observed for our Superflares.  Following the argument of Shibayama, this is good evidence that the Superflares on V2487 Oph result from magnetic reconnection.

\subsection{Time Between Superflares}

In Section 4.1, we found that the flare energy, $E$, is proportional to $\Delta T$.  This relation is the hallmark of any mechanism for energy release where some reservoir of energy is being filled, only to have it reach some critical threshold, whereupon the energy is dumped quickly, appearing as a flare, then to have the reservoir refilled again until it reaches the critical threshold for the next flare.  If a large amount of energy is dumped in one event, then the reservoir is relatively empty and takes a long time to refill to the critical threshold.  If only a small amount of energy is dumped in one event, then the reservoir is already near the critical threshold, so the next flare will occur soon.  That is, $\Delta T$$\propto$$E$, just as seen in the V2487 Oph Superflares.

In the context of the gravitational energy scenario, the reservoir being steadily filled is a pool of gas waiting to be dumped when some trigger threshold is met.  For the case of a dwarf nova event, the energy reservoir would be the gas circulating in the disc, and the trigger is the well-known accretion instability.  However, DN events do {\it not} show any correlation between the total outburst energy and the waiting time until the outburst.  (We have confirmed this with our analysis of 650 dwarf nova events from the century-long AAVSO light curve of the prototype SS Cyg.)  For the case of Type {\rm II} X-ray bursts, we have an example of an instability in the deep inner disc, albeit with this being around a neutron star.  These events do have a moderately tight correlation between the burst energy and the time until the next burst, where an energetic burst depletes most of the gas in the inner disc, resulting in a relatively long time to fill the inner disc back to the trigger level for another dump (Lewin et al. 1993).  This would appear to be a precedent that accretion from an inner disc can produce a proportionality between $E$ and $\Delta T$.  However, the instabilities that work for the Type {\rm II} bursts (e.g., the Lightman--Eardley instability) require either the very deep gravity well of a neutron star or its incredibly high magnetic field, such that they could not work around a white dwarf (Lewin et al. 1993; Frank et al. 2002).  So we do not know of any physical mechanism that could produce accretion instabilities in the inner region of a disc around a white dwarf.  In all, we are left with no understanding of how a gravitational energy scenario can produce a linear relation between $E$ and $\Delta T$.

In the context of the nuclear burning scenario, the reservoir being steadily filled is the mass of accreted material on the surface of the white dwarf, and the trigger criterion is that the flare starts when the accreted layer accumulates the ignition mass.  For the case of nova eruptions, we can only measure $\Delta T$ for RNe.  For the RNe for which we have three or more eruption with confidence that no eruptions were missed, the observed $\Delta T$ values change up up to a factor of 3.7$\times$ for T Pyx, 1.8$\times$ for U Sco, and 3.1$\times$ for RS Oph. (Schaefer 2010).  Despite these large variations, the light curves remain identical, so the $E$ value is sensibly a constant (Schaefer 2010).  Likely, the variations are a result of variable accretion rates making for unsteady filling of the reservoir of material accumulated on the white dwarf surface (Schaefer 2005).  For the case of Type {\rm I} X-ray bursts, it is encouraging that these display the expected energy versus waiting time relation (Lewin et al. 1993). 

In the context of the magnetic reconnection scenario, we know of no ready means to calculate $E$ or $\Delta T$, so we can only look to known examples of stellar magnetic reconnection; solar flares, flare stars, and Superflare stars.  The idea is that the reservoir of energy is the magnetic field over the accretion disc, with the refilling caused by magnetic stress resulting from the winding-up of field lines attached to either the disc or companion.  When the magnetic stress rises to some critical level, reconnection breaks down the field lines.  With such a set up, there should be a proportionality relation between $E$ and $\Delta T$.  Importantly, the proportionality can be seen only when there is {\it one} reservoir of energy, because if a series of events from two-or-more reservoirs intermix then the $\Delta T$ values are all mixed up.  Further, the filling of the reservoirs must be steady for the proportionality to work.  For the case of white light solar flares, the flares are produced over multiple sunspot regions, with the twisting of field lines being unsteady in time, so we neither expect nor observe any $E$$\propto$$\Delta T$ relation.  Similarly for the case of ordinary main-sequence flare stars, the flares come from multiple reservoirs, as demonstrated by many {\it Kepler} light curves that show successive flares that differ in rotational phase by near 180$\degr$.  That is, successive flares often come from different spots in different hemispheres.  So, for flare stars, we neither expect nor observe any $E$$\propto$$\Delta T$ relation.  Similarly for the case of Superflares on solar-like stars, as well-observed with the {\it Kepler} mission, successive Superflares are often caused by separate regions in opposite hemispheres (Maehara et al. 2012), so more than one energy reservoir is needed, so we neither expect nor observe any $E$$\propto$$\Delta T$ relation.  So we do not have any precedents for stellar magnetic reconnection flares where only {\it one} energy reservoir is involved.  The only information that we are left with is the strong theoretical expectation that the V2487 Oph Superflares should follow the $E$$\propto$$\Delta T$ relation if only one reservoir of magnetic energy is involved.  Indeed, we can turn this around to get the insight that V2487 Oph follows the $E$$\propto$$\Delta T$ relation, albeit with substantial scatter, therefore only one reservoir is involved.  For example, magnetic loops with a footprint in the accretion disc will rapidly be mixed up by differential rotation and turbulence in the disc's gas, so the field lines would result in one overall structure.

In summary, all three energy sources (gravitational, nuclear, and magnetic) can apparently explain the $\Delta T$$\propto$$E$ case for V2487 Oph, none of these explanations is fully satisfactory.

\subsection{Superflare Rise Times}

The rise times ($\tau_{rise}$) represent information that can be compared to physical limits, perhaps constraining the possible physics of the flare mechanism. Many of the Superflares have sharp initial rises with over half the peak flux coming in less than the 59-second time resolution.  So what energy sources and scenarios have mechanisms that give rise times $\lesssim$59 seconds?

The rise time within the gravitational energy source depends on the scenario.  The rise times for events involving instabilities in the middle or outer part of the accretion disc (i.e., DN events) have time-scales of a day or so, while the rise times cannot be faster than the free fall time, which is similar to the orbital period.  So in the gravitational energy scenario, we can get fast rise times only when the inner disc is involved.  The orbital period just above the surface of a 1.35 M$_{sun}$ white dwarf is near 1.6 seconds.   An empirical demonstration of fast rise times from accretion instabilities in the inner edge of the accretion disc is that Type {\rm II} X-ray bursts have rise times of order 1 second, albeit around a neutron star (Lewin, van Paradijs, \& Taam 1993).  So these scenarios can reproduce the observed fast rise times in V2487 Oph Superflares.

The rise time within the nuclear burning scenario is a combination of time-scales from the ignition of the fuel, the energy transfer to the surface of the accreted layer, and the horizontal propagation across the region with fuel.  The dynamical time-scale, the nuclear burning time-scale, and the convective turn-over time-scale are all on the order of seconds near the peak of the runaway (Starrfield, Iliadis, \& Hix 2008).  The horizontal propagation of the flame will go at a velocity between 0.3 and 10,000 m s$^{-1}$, depending on the conditions of convection and turbulence, with a best estimate of 140 m s$^{-1}$ (Fryxell \& Woosley 1982).  For a plausible spot size of 400 m in radius, the flame will start the whole region burning with a rise time of order 3 seconds.  So the thermonuclear runaway idea can easily produce the fast observed Superflare rise times.

The rise time within the magnetic reconnection scenario is best seen from the observed rise times exhibited by known stellar magnetic reconnection events.  White light solar flares (WLSFs)
 have typical rise times ranging from under one minute to roughly ten minutes.  Indeed, the all-time brightest white light solar flare has a reported rise time of greatly less than 60 seconds (Carrington 1859).  The typical rise times for flare stars ranges from 10 seconds to five minutes.  Typical flares on Superflare stars have rise times that range from under one minute up to roughly ten minutes.  So the magnetic reconnection scenario readily has the fast rise times as seen for V2487 Oph.

In summary, all these scenarios can easily explain that the V2487 Oph rise times can be faster than 59 seconds.

\subsection{Superflare Light Curve Shapes}

The shape of the light curves (LCs) of the V2487 Oph Superflares appears to be universal, with an initial spike followed by a tail.  There can be 1--3 initial spikes, the tail can have various bumps and FRED maxima, while the relative brightness of the spike and tail can vary substantially.  Still, the universality of the initial impulsive spike followed by a tail should be diagnostic.

The light curve shape within the gravitational energy scenario can be learned from various examples.  Events caused by an instability in the disc dumping mass to the inner disc (i.e., dwarf novae) have a rounded and nearly-symmetrical light curve shape, with no fast rise or exponential tail.  Events caused by instabilities in the inner disc around a neutron star (i.e., Type {\rm II} X-ray bursts) vary widely in shape, but typically have rounded profiles with no initial spike and no emission tail.

The expected light curve for thermonuclear burning on a small spot will just be the light curve for the surface luminosity for a 1.35 M$_{\odot}$ white dwarf, as displayed in Starrfield et al. (2008).  These show a very fast rise of under a minute or so, a fast spike near the top of the flare, and an exponential decay with a time-scale around two hours.  For some libraries of nuclear reaction rates, the calculated light curves even show a spike followed by a FRED.  The theory expectations are clear, yet we can test their generality by looking at other cases of thermonuclear burning on a collapsed star.  For novae, the light curve shapes can be characterized as starting off with an initial spike, followed by an `exponential' tail with various bumps.  For Type {\rm I} X-ray bursts, the X-ray light curve shapes (see Lewin et al. 1993) that are indistinguishable from V2487 Oph Superflare light curve shapes.  So the nuclear burning scenario can produce light curves with shapes and durations similar to those of the observed Superflares.

The expected light curve for magnetic reconnection flares cannot be calculated in general for the unknown case of V2487 Oph, so all we can do is look at light curve shapes from different cases of stellar magnetic reconnection.  White light solar flares have a substantial fraction of the flare energy in the optical, while the {\it average} light curve of the brightest flares has the highest peak being a symmetric spike with a duration of 15 minutes (Kretzschmar 2011).  This initial spike in solar flares is even given a special name, as the `impulsive phase'.  Individual white light flares can have fast rises and have a bumpy and spiky tail trailing off from the highest peak.  Flare stars have light curves with fast rises to a spike, with an `exponential' tail, plus occasional plateaus (like FRED peaks) and spikes just after the initial impulsive spike.  Superflares on Superflare stars have light curve shapes that characteristically have fast rises for initial spikes, `exponential' tails, plus occasional extra spikes and FRED shaped plateaus (Maehara et al. 2012; Balona 2012; Shibayama et al. 2013).  And 18 per cent of Superflares show small-amplitude bumps in the declining tail, and a magneto-hydrodynamic force acting on the field loops has been proposed as the mechanism (Balona et al. 2015).  In all, stellar magnetic reconnection events produce light curve shapes that look like the Superflares on V2487 Oph.

In summary, gravitational energy events do not produce light curve shapes like the Superflares in V2487 Oph, whereas both nuclear burning events and magnetic reconnection events produce light curves shapes that are stunningly like those seen in V2487 Oph.

\subsection{Superflare Durations}

The V2487 Oph Superflares have the durations scaling as a power law of the burst energy, with $D_{eq}$$\propto$$E^{0.44\pm0.03}$.  This distinct relation should be characteristic of the flare mechanism.

Durations for the gravitational energy scenario, and their relation to flare energies, cannot be calculated for any general case, and certainly not when the mechanism for accretion instability in the inner disc is not known.  But we can look at two cases where gravitational instabilities are dumping mass on to a collapsed star to create flares; for dwarf novae and for Type {\rm II} X-ray bursts from the Rapid Burster.  For dwarf novae, we have taken the energies and durations of 704 outbursts from the prototype SS Cyg from the AAVSO century-long light curve.  We find a fairly tight relation between eruption duration and energy, with the durations going as $E^{0.598\pm0.004}$.  While this is at least a power law, the slope is greatly different from that of V2487 Oph.  Detailed analysis of data for individual bursts gives power law slopes variously from +1.0 to +1.6 (Kunieda et al. 1984; Stella et al. 1988; Lewin et al. 1993).  The point is that Type {\rm II} X-ray bursts display a duration/energy relation that is far from that observed for V2487 Oph.

Durations for the nuclear burning scenario are controlled by the local physics (reaction rates, convection time-scales, cooling rates, and so on), which is independent from the energy production.  For nova eruptions, thermonuclear runaways involving the entire surface of a white dwarf, the only cases where we have multiple eruptions from the same conditions are for the RNe, including V2487 Oph itself.  In all eruptions from each RNe, the light curves are essentially identical in peak magnitudes, durations, and light curve shapes (Schaefer 2010).  This is simply because the important conditions ($M_{WD}$, $\dot{M}$, and the gas composition) remain the same, while the triggering conditions ensure that the trigger masses are identical.  Still, the duration is a constant from event to event in this case powered by nuclear burning.  For Type {\rm I} X-ray bursts, most have a linear relation between the flare energy and the peak flux (Lewin et al. 1993), which is to say that the duration is roughly a constant, which is to say that the durations scale as $E^{0.0}$.  We can generalize with added insight.  Let us consider some region with surface area $A$ that participates in the runaway.  The time-scale of energy production at the surface of $A$ is determined by the local conditions and physics (Starrfield et al. 2008), so the light coming from this small area has a duration $D_A$ and an energy $E_A$.  If the burning region is, say, 10$\times$A in area, then all the areas will start their burning sequence nearly simultaneously due to the fast lateral propagation of the flame, and each area will burn out around the same time so that the overall flare has the same duration $D_A$ independent of area.  But with 10$\times$A burning, the total flare energy will be 10$\times$$E_A$.  So for all nuclear burning scenarios, we expect that $D_A$$\propto$$E^{0.0}$ (i.e. the duration does not change with energy), in strong violation of the observed case for V2487 Oph. 

Durations for stellar magnetic reconnection events have been reported in a variety of papers.  Namekata et al. (2017) report that white light solar flares have $D$$\propto$$E^{0.38\pm0.06}$.  For ordinary flare stars (M dwarfs), our analysis of the data listings in Yang \& Liu (2019) shows 
that individual flare stars vary with $D$$\propto$$E^{0.47\pm0.07}$.  Superflares on solar-like stars have $D$$\propto$$E^{0.39\pm0.03}$ as reported by Maehara et al. (2015) for {\it Kepler} data, and have  $D$$\propto$$E^{0.42\pm0.01}$ for {\it TESS} data as reported by Tu et al. (2020).  In all, it appears to be a ubiquitous relation amongst all stellar reconnection events that the durations vary close to $E^{+0.42}$.  We see this as a good match with the Superflares of V2487 Oph scaling as $D_{eq}$$\propto$$E^{0.44\pm0.03}$.

In summary, neither gravitational energy nor nuclear burning can produce the observed duration distributions, whereas the universal distribution observed for stellar magnetic reconnection events matches closely that observed for V2487 Oph.  This result is so distinct that we take this to be strong evidence that the V2487 Oph Superflares are caused by magnetic reconnection.

\subsection{Summary}

For the gravitational energy scenarios, we can rule out packets of gas hitting the hotspot at the edge of the accretion disc and we can rule out instabilities in the middle or outer disc, as these do not reproduce the energetics requirements, the observed fast rise times, nor the flare light curve shapes.  But these constraints can be satisfied if we move the instability to the inner accretion disc.  Type {\rm II} X-ray bursts are not useful precedents because all of the many conceived instabilities in the innermost depths of the accretion disc all require the extreme conditions found only close to a neutron star (Lewin et al. 1993; Frank et al. 2002), so any such proposed mechanism does not apply to V2487 Oph.   Even in the inner disc, these scenarios apparently cannot explain the ($dN$/$dE$)$\propto$$E^{-2.34 \pm 0.35}$, the $D_{eq}$$\propto$$E^{0.44\pm0.03}$ relations, nor the light curve shapes.  Taken together, we view these as effective refutations of the gravitational energy scenario.

The nuclear burning scenario can account well for the available energy, the fast rise time, the light curve shape, the typical duration (but not the relation between duration and energy), and maybe the $E$$\propto$$\Delta T$ relation.  This is sounding pretty good.  To get this to work, we are forced to have the nuclear burning occurring only inside {\it one} small region on the surface of the white dwarf, presumably with just $\sim$0.0001 of the accreting material funneled on to a patch only $\sim$400 m in size.  Thus, thermonuclear runaway burning on a small region on the surface of the white dwarf is expected to produce a flare that looks nearly the same as our observed Superflares on V2487 Oph.  This non-standard requirement is presumably met by having a high-order magnetic field configuration funneling the incoming gas to the region, with this requirement seeming to be plausible since V2487 Oph is likely a magnetic system (Hernanz \& Sala 2002).  We view the thermonuclear runaway idea to be hopeful, because it predicts Superflares similar to those observed and because the situation is familiar with well-known physics.  Nevertheless, the scenario cannot account for the observed ($dN$/$dE$)$\propto$$E^{-2.34 \pm 0.35}$ and $D_{eq}$$\propto$$E^{0.44\pm0.03}$ relations.  These two relations are distinctive and diagnostic, so the failure to match these observations is critical.  With these failures to be able to match these two well-measured relations, the nuclear burning scenario can only be taken as `refuted'.

For the magnetic reconnection scenario, we see that the V2487 Oph Superflares look and act like known Superflares on stars just like the V2487 Oph companion star.  In particular, the many Superflares on G-type main sequence and sub-giant stars are indistinguishable from those on V2487 Oph in terms of their rise times, durations, light curve shapes, and energies.  Importantly, Superflares on V2487 Oph follow the ($dN$/$dE$)$\propto$$E^{-2.34 \pm 0.35}$ and $D_{eq}$$\propto$$E^{0.44\pm0.03}$ relations, exactly as predicted for all magnetic reconnection events.  Taken together, these relations are unique to Superflares, and we view all this as proof that the V2487 Oph Superflares are caused by magnetic reconnection.

\section{Conclusions}

V2487 Oph is important as one of only ten known RNe in our Milky Way, and it has the poorest observational record.  We have presented large amounts of photometry, three independent data sets of which are pointing to an orbital period of 1.24$\pm$0.02 days.

The CV evolution issues of V2487 Oph are now overshadowed by our discovery that the system displays Superflares.  They are given the descriptive name `Superflares' due to their tremendous energy, over 10$^{38}$ ergs.  The Superflares occur roughly daily, with a total energy of 10$^{41}$ ergs per year.  The light curve shapes invariably start with an impulsive sharp spike, and are always followed by some sort of an `exponential' tail, that often has further spikes and various small bumps.  Their rise times are often shorter than 59 seconds, and their durations are typically one hour.  The flare amplitude rises up to 1.10 mag.  The Superflares are isolated in time, with no apparent activity between Superflares.  V2487 Oph is seen to have Superflares from 2016 on-going until late 2021.

The Superflares obey three relations.  First, the distribution of flare energies follows a power law such that ($dN$/$dE$)$\propto$$E^{-2.34 \pm 0.35}$.  This rules out mechanisms that produce either a roughly constant flare energy, or mechanisms that have some cut off in energy.  Second, the flare energy is proportional to the time to the next burst, $E$$\propto$$\Delta T$.  This second relation is clearly pointing to the physical mechanism for the flares involving {\it one} reservoir of energy being steadily filled until it reaches some threshold, whereupon it dumps much of the energy which appears as a flare, only to have the reservoir refill and then start another cycle.  Third, the flare durations scale as a power law of the flare energy, $D_{eq}$$\propto$$E^{0.44\pm0.03}$.  This rules out the various ideas where the mechanism produces flares with some capped luminosity (say, by the Eddington limit), and it rules out ideas where the duration is independent of the energy.

The Superflares are startling and exciting.  This is the first time that any such phenomenon has been seen from any CV or any interacting binary.  The big question then becomes the nature of the Superflares.  

In this paper we explored three plausible energy sources: gravitational, nuclear, and magnetic.  The gravitational energy scenario has some sort of an accretion instability, presumably in the inner region of the accretion disc, such that gas is speedily dumped to a lower position with the potential energy going to thermal and then radiative energy, to appear as a Superflare.  The nuclear energy scenario has accreting gases accumulating on some small region on the surface of the white dwarf, reaching the usual critical mass to initiate runaway thermonuclear burning, and the hot gases produce the light of the Superflare.  The magnetic energy scenario has loops of strong magnetic fields twisted and amplified by the motions of the loop's footprint (in either the disc or the companion star), increased to the point where the magnetic stresses reach a critical threshold, causing reconnections of the field lines to lower energy configurations, whereupon the resultant Alfven waves accelerate ambient electrons to very high energies, with the electrons slamming into their footprints, with this electron energy speedily getting converted in part into the radiant energy of the Superflare.  All three of these energy sources have greatly more than enough energy to power the Superflares.

The gravitational energy and nuclear energy scenarios have problems.  The accretion instability idea does not work out energetically when dealing with the outer or middle disc.  Accretion instability in the inner disc has the deep problem that no one has devised a mechanism that would work around a white dwarf.  Still, these gravitational energy ideas apparently cannot reproduce any of the three key relations.  So the gravitational energy scenario appears to be refuted.  The nuclear burning energy source idea only works if a complex magnetic field on the white dwarf funnels down $\sim$0.0001 of the accretion flow on to a small region only $\sim$400 m in size.  This `nuclear' hypothesis has the strong advantage that it is expected to produce flares with similar rise times, durations, and light curve shapes as are seen for V2487 Oph.  The problems for the nuclear burning idea is that it apparently cannot account for the two relations with ($dN$/$dE$)$\propto$$E^{-2.34 \pm 0.35}$ and $D_{eq}$$\propto$$E^{0.44\pm0.03}$.  This failure refutes the nuclear energy scenario.

The magnetic energy scenario fits perfectly.  The big advantage is that we already know that Superflares occur on ordinary G-type stars (just like the companion star in V2487 Oph) with these being indistinguishable in terms of energies, rise times, durations, and light curve shapes.  Further and critically, Superflares on both solar-like stars and V2487 Oph obey the relations, with ($dN$/$dE$)$\propto$$E^{-2.34 \pm 0.35}$ and $D_{eq}$$\propto$$E^{0.44\pm0.03}$.  Indeed, these relations have been previously identified as effectively universal laws for stellar magnetic reconnection events.  So in conclusion, with such a perfect match between Superflares on solar-like stars and Superflares on V2487 Oph, we can be relatively confident that the Superflares are caused by magnetic reconnection within the binary system.

So V2487 Oph is a Superflare star, with the Superflares powered by large-scale magnetic reconnection.  Indeed, V2487 Oph is the most extreme Superflare star, in terms of its energy.  V2487 Oph has $E$ up to 1.6$\times$10$^{39}$ ergs, which is over a thousand times larger than the highest known Superflare from any ordinary solar-like star, as seen by {\it Kepler} or {\it TESS}, at 10$^{36}$ erg ( Maehara et al. 2012; Shibayama et al. 2015).  Even for the most extreme known Superflare  (the triply independently-confirmed 1899 event on solar-analog S For that had energy $\sim$2$\times$10$^{38}$ ergs, see Schaefer et al. 2000) has lower $E$, and these extreme Superflares are very rare, perhaps to the level of once-per-century.  For lower energy systems, even the fastest Superflare stars erupt only once every ten days on average for $E$$>$10$^{34}$ ergs (Shibayama et al. 2013), for a yearly energy budget of roughly 10$^{36}$ ergs.   So, not only is V2487 Oph with the most energetic observed Superflare, but also has the highest known Superflare frequency.  The yearly average energy budget for V2487 Oph Superflares is many orders of magnitude larger than the energy budget for any other Superflare star.  So maybe we should be labelling V2487 Oph as a `Super-Superflare star'.

Why is V2487 Oph such an extreme Superflare star?  Well, the RN is unique amongst known Superflare stars in having an accretion disc around a collapsed star.  Presumably, V2487 Oph gets its powers boosted by the disc or the white dwarf.  It is easy to imagine that the very high magnetic fields around white dwarfs make for greatly higher $E$ than for field stars.  But this raises the question of why other novae, CVs, and X-ray binaries do not have Superflares?  For example, the presumably identical RN U Sco has extensive photometric time series, yet not a hint of any flares of any amplitude.  Large numbers of novae, CVs, and X-ray binaries have been closely monitored, with no Superflares, so the events must either be very rare on any one system, or the fraction of CVs with frequent Superflares must be less than something like 1 per cent.  In either case, V2487 Oph must have some rare property that makes its events both frequent and high-energy.

Even with the incredible 67 days of a nearly continuous light curve from {\rm Kepler}, V2487 Oph remains a poorly observed system.  Likely, the biggest observational need at present is to get long runs of spectroscopy in quiescence, which would test the orbital period and might yield spectra of a Superflare.  With magnetic reconnection events producing power law SEDs from radio to X-ray, it would be interesting to look at V2487 Oph in radio, infrared, and X-ray bands.  And we should remember that for the likely $\sim$18 years recurrence time, V2487 Oph should be having a third nova event any year now, so both professional and amateur observers should be regularly monitoring the system.

The V2487 Oph Superflares are from large-scale magnetic reconnection, but this leaves various critical high-level questions unanswered.  Likely the most important is to determine the configuration of the magnetic field lines.   Where are the footprints of the magnetic loops, and what is the loop length and cross section?  What is the magnetic field strength?  There is a strong need for realistic and detailed models for the amplification and twisting of the field lines, plus some physics criterion for knowing when the field lines will reconnect.  (The need for better micro-physics theory actually applies to reconnection events with solar flares, ordinary flare stars, and for Superflare stars all across the HR diagram.)  And we need to know why the many other CVs, especially the `sister-RN' U Sco, do not have any known Superflares despite very intensive observations over many decades.  And we need some sort of understanding as to why V2487 Oph is so extreme amongst all other Superflare stars.

\section{Acknowledgements}

We thank the allocation committees at McDonald Observatory, CTIO, and the {\it Kepler K2} mission for granting us time to observe V2487 Oph.  We thank the many AAVSO observers who tirelessly produced professional quality photometry, despite no pay and no specific knowledge of how their photometry will be used.  We thank the PTF, Pan-STARRS, and ZTF teams for producing their light curves, with the ZTF flare providing a strong proof that the Superflares are intrinsic to the old nova system. 

This paper includes data collected by the Kepler mission and obtained from the MAST data archive at the Space Telescope Science Institute (STScI). Funding for the Kepler mission is provided by the NASA Science Mission Directorate. STScI is operated by the Association of Universities for Research in Astronomy, Inc., under NASA contract NAS 5-26555.  This research made use of Astropy \citep{2018AJ....156..123A}, Lightkurve, a Python package for Kepler and TESS data analysis \citep{2018ascl.soft12013L}, and EVEREST \citep{2016AJ....152..100L, 2018AJ....156...99L}.

Partial financial support for BES and AP's early work was provided by the National Science Foundation, under grant AST-0708079. Partial financial support for AP was provided by the Oak Ridge National Laboratory's Ralph E. Powe Junior Faculty Enhancement Award. Financial support for SZ was provided by the College of Charleston's School of Science and Mathematics Summer Research Award. AP would additionally like to thank the LWS Slack Channel and MS in particular for many useful co-writing sessions.

\section{Data Availability}

All of the photometry data used in this paper are explicitly given in Tables 2, 3, and 4, and are publicly available from the references and links in Table 1.


{}

\bsp	
\label{lastpage}
\end{document}